\newif\ifraw   
\rawfalse
\documentclass[11pt,fleqn,leqno]{article}
\usepackage{amsmath}
\usepackage{amssymb}
\usepackage{amsthm}
\catcode`\@=11
\catcode`\@=12
\usepackage{enumerate}
\usepackage{xspace}
\newcommand{\zeile}{\hfill\\}
\newcommand{\komma}{\quad ,}
\newcommand{\punkt}{\quad .}
\swapnumbers
\theoremstyle{plain}
\newtheorem{theorem}{Theorem}[section]
\newtheorem*{theorem*}{Theorem}
\newtheorem{proposition}[theorem]{Proposition}
\newtheorem*{proposition*}{Proposition}

\newtheorem*{lemma*}{Lemma}
\newtheorem{corollary}[theorem]{Corollary}
\newtheorem*{corollary*}{Corollary}
\theoremstyle{definition}
\newtheorem{definition}[theorem]{Definition}
\newtheorem*{definition*}{Definition}
\theoremstyle{remark}

\newtheorem*{remark*}{Remark}

\newtheorem*{note*}{Note}

\newtheorem*{cit*}{Citation}
\newcommand{\ein}{\hangindent1cm\hangafter1\zeile}

\newcommand{\RR}{\ensuremath{\mathbb{R}}}   
\newcommand{\ZZ}{\ensuremath{\mathbb{Z}}}   
\renewcommand{\d}{\partial} 

\newcommand{\abs}[1]{\lvert#1\rvert}

\newcommand{\lnorm}[2]{\lVert#1\rVert_{L^{#2}}}

\DeclareMathOperator{\tr}{tr}
\renewcommand{\div}{\ensuremath{\text{div}}\,}
\DeclareMathOperator{\grad}{grad}

\DeclareMathOperator{\id}{id}

\DeclareMathOperator{\vol}{Vol}

\DeclareMathOperator{\supp}{supp}
\DeclareMathOperator{\Gl}{GL}

\newcommand{\leftsuper}[1]{{}^{\scriptscriptstyle #1}}

\newcommand{\viergamma}{\leftsuper{4}\gamma}
\newcommand{\viernabla}{\leftsuper{4}\nabla}
\newcommand{\vierR}{\leftsuper{4}\!R}
\newcommand{\viere}{\leftsuper{4}\!e}
\newcommand{\viersigma}{\leftsuper{4}\!\sigma}

\newcommand{\obar}[1]{\smash[t]{\overset{\rule[-0.5pt]{1ex}{0.5pt}}{#1}}}
\newcommand{\ubar}[1]{\smash[b]{\underset{\rule[5pt]{1ex}{0.5pt}}{#1}}}
\newcommand{\mass}{\ensuremath{\text{\sc m}}\xspace}
\ifraw
  \input wiggly.tex  
\fi
\title{Global Prescribed Mean Curvature foliations in cosmological
  spacetimes with matter\\ Part II}
\author{Oliver Henkel\thanks{Present address: Heinrich--Hertz--Institut
    f\"ur Nachrichtentechnik Berlin GmbH, Einsteinufer 37, 10587 Berlin,
    Germany}\\ 
  Max Planck Institute for Gravitational Physics\\ Am M\"uhlenberg 1\\
  14476 Golm, Germany}  
\date{October 18, 2001}
\begin{document}
\maketitle
\begin{abstract}
   This second part is devoted to the investigation of global properties of
   Prescribed Mean Curvature (PMC) foliations in cosmological spacetimes with
   local $U(1) \times U(1)$ symmetry and matter described by the Vlasov
   equation.
   It turns out, that these spacetimes admit a global foliation by
   PMC surfaces, as well, but the techniques to achieve this goal are more
   complex than in the cases considered in part I. 
\end{abstract}
\newpage
\tableofcontents
%
\ifraw
\renewcommand{\emph}{\underline}
\renewcommand{\textbf}{\underwiggle} 
\renewcommand{\baselinestretch}{2}
\tt
\fi
\section{Introduction}
For the main motivation see the introduction in part I, \cite{h3}. Only
some remarks, special to the present situation remain.\zeile
The structure of this second part is as close as possible to the
structure of part one. Although the preliminary section 2 of part I has
been omitted, the formulas here refer to it as well. 

The spacetimes considered here are cosmological spacetimes with two
commuting local Killing vector fields. This symmetry will be referred to as
(local) $U(1) \times U(1)$ symmetry. In comparison with part I, there are
now only two (local) Killing fields instead of three, generalizing the
plane symmetric case of \cite{h3}. The
absence of a third Killing field requires a more detailed description of
the geometry and a deeper analysis to control the
momenta of the Vlasov particles. Thus the estimates in this work rely on the
simple structure of the Vlasov equation. Despite this there seems to
exist no crucial obstructions for other types of 'well behaved' matter.
%
%
%
%
%


%
%
%
%
%
%
%
\section{Spacetimes with local $U(1) \times U(1)$ symmetry}
\label{s.gowdysym} 
\subsection{The geometry of spacetimes with local 
  $U(1) \times U(1)$ symmetry} 
Following the analysis in \cite{r}, we consider now the globally
hyperbolic spacetime $(M,g)$ with
topology $\RR \times \Sigma$, where $\Sigma$ denotes a bundle with
base $S^1$ and compact orientable fibre $F$. As usual it is assumed,
that the submanifolds $\{t\} \times \Sigma$ are Cauchy hypersurfaces in
$M$. The coordinates of $\Sigma$ are denoted by $(x,y^2,y^3)=(x,y^A)$, 
$A=2,3$, where $(y^A)$ denote coordinates on $F$. As usual, Greek
indices range in the interval $0,\dots,3$, 
lower case Latin indices from the middle of the alphabet take values
in $1,\dots,3$, while those from the beginning of the alphabet take
the values $0,1$ and upper case ones are confined to the values
$2,3$.\zeile 
The covering map $\RR \longrightarrow S^1$ defines a pullback of the
bundle $\Sigma$ with base $\RR$, hence we get a trivial bundle. If
$\hat{F}$ denotes the universal covering space of $F$, we get a 
natural covering 
$\hat{\Sigma}=\RR \times \hat{F}$ of $\Sigma$ with canonical
projection $p$. Now we can associate a spacetime
$(\hat{M},\hat{g})$, where 
$\hat{M}=\RR \times \hat{\Sigma}$ and $\hat{g}$ is the pullback
of $g$ under the projection $\id \times p$.

The fibres $\hat{F}$ in the trivial bundle $\hat{M}$ are assumed
to be the orbits of a two dimensional translation group $G$ of isometries of
$\hat{g}$. Hence $\hat{F}$ is the Euclidean space form $E_2$ and
$F=E_2/\Gamma$, for a discrete subgroup $\Gamma$ of $G$. 
The compactness and orientability of $F$ implies then $F=T^2$, so 
$\Gamma$ can be represented by a two-parameter lattice, $\Sigma$
turns out to be a torus bundle over the circle and
the induced action of $G$ on $(M,g)$ is given by the quotient action 
$G/\Gamma = U(1) \times U(1)$ with orbits $F$. As in \cite{h3}
we will call the orbits 
surfaces of symmetry, hypersurfaces in $M$ diffeomorphic to 
$\{t\} \times \Sigma$, which consist of a union of surfaces of
symmetry will be called symmetric surfaces and we call $(M,g)$ 
a spacetime with local $U(1) \times U(1)$ symmetry.\zeile
The induced action of $U(1) \times U(1)$ on $(M,g)$
is local, if the bundle is non-trivial. In that case  
we have to deal with non-trivial transformations for the transition 
$x \longmapsto x+2\pi$ in $S^1$ by lattice preserving translations and
automorphisms $\Gl(2,\ZZ)$ of $G$. To 
represent the metric $\obar{g}$ of the orbits in $M$ at first locally for
$x \in [0,2\pi[$, define the area 
radius by $r := \sqrt{(4\pi)^{-1}\vol(F)}$, and write the metric as
\begin{equation}\label{e.gobar2}
   \obar{g} = r^2 \tilde{g}
   \komma
\end{equation}
with a metric $\tilde{g}$ of unit determinant. Due to symmetry both
quantities, $r$ and $\tilde{g}$ do not depend on the points of $F$,
and one easily verifies, that the curvature of $\obar{g}$ vanishes, as
required. For 
$\tilde{g}$ there are two remaining degrees of freedom, $V$ and $W$, and 
we can use them to parametrize the metric:
\begin{equation}\label{e.gtilde2}
   \tilde{g} = 
      \begin{pmatrix} 
         e^W \cosh V & \sinh V \\
         \sinh V     & e^{-W} \cosh V
      \end{pmatrix}
   \punkt
\end{equation}
If the bundle is trivial, then this representation of the metric is
global, but if the bundle is not trivial, then the translation of
$2\pi$ in $S^1$ induces a transformation of $\obar{g}$ by
an element $Z$ of $\Gl(2,\ZZ)$,  
\begin{equation}\label{e.gobartransform}
   \obar{g}(x+2\pi) = Z^T\obar{g}(x)Z
   \komma
\end{equation}
which fixes the geometry of the spacetime.

Given a globally acting symmetry we can specialize to some well known
geometries: 
If $V=W=0$ then we get the plane symmetric case $\epsilon=0$ of
\cite{h3}. Setting only $V=0$ we get a symmetry called
polarized, corresponding to the reflection symmetries 
$y^2 \longmapsto -y^2$ or $y^3 \longmapsto -y^3$ respectively. If the
composition of 
these reflections is an isometry (regardless whether the individual
reflections act isometrically), we call this symmetry to be of
\emph{Gowdy--type},
since the Gowdy spacetimes are defined by this symmetry and the
additional requirement that the spacetime is vacuum. Thus, in general 
$V \neq 0$ in a spacetime with Gowdy-type symmetry, but if $V$ vanishes
also we call the Gowdy-type symmetry polarized.

We construct the coordinate system $\{x^{\mu}\}$ mentioned in the
beginning of this section locally by the
following procedure. Consider first an arbitrary symmetric Cauchy
surface $S$ in $M$. Then we extend the coordinates $(y^A)$ of $F$ to
a Gaussian neighbourhood of $F$ in $S$. Later on we will do some rescaling
along $S^1$, such that we choose the coordinates, such that the metric 
$h$ of $S$ takes the general form $h=A^2\,dx + \obar{g}$. Now we embed 
this structure into the spacetime. In general the time coordinate $t$
defines lapse $N$ and shift $\nu=\nu^i\,\d_i$, thus the spacetime
metric $g$ is not in block diagonal form. The components $g_{0A}$
which prevent $g$ being block diagonal can 
be represented by $\obar{\nu}_A:=\obar{g}_{AB}\,\nu^B$. 
\subsubsection{The 2+2--geometry}\label{ss.2+2}
Now we want to investigate the geometry of $(F,\obar{g})$ in $(M,g)$
more closely. As before let $S$ denote a symmetric Cauchy surface in $M$,
foliated by its symmetric surfaces. The unit normal of $F$ in 
$S \subset M$ will be denoted by $m$ and $n$ is the unit normal of $S$ in
$M$, as usual. Then there are some canonical geometrical objects induced:
\begin{enumerate}[(a)]
\item the two second fundamental forms
   \begin{subequations}\label{e.lambdakappa2}
      \begin{gather}
         \lambda(v,w) = g(\viernabla_v w,m) \label{e.lambda2} \\
         \kappa(v,w) = g(\viernabla_v w,n) \label{e.kappa2} 
         \komma
      \end{gather}
   \end{subequations}
   for $v,w \in TF$ with arbitrary extensions to vector fields, in order to
   make the expressions well defined. In the sequel it turns out to be 
   convenient, to deal with the trace free parts $\tilde{\lambda}$ and
   $\tilde{\kappa}$ instead, defined by 
   $\lambda = \tilde{\lambda} + \tfrac{1}{2}\tr\lambda\,\obar{g}$ and
   $\kappa = \tilde{\kappa} + \tfrac{1}{2}\tr\kappa\,\obar{g}$
   respectively. In the given coordinates we have explicit formulas:
   \begin{gather*}
      \lambda_{AB} = -\tfrac{1}{2}m(\obar{g}_{AB}) 
                   = -\tfrac{1}{2} A^{-1}(\obar{g}_{AB})' \\
      \tr\lambda = -\tfrac{1}{2}\obar{g}^{AB}\,m(\obar{g}_{AB})
                 = -\tfrac{1}{2}\frac{m(\det \obar{g})}{\det \obar{g}}
                 = -\tfrac{2}{r}m(r) = -2A^{-2}A' \\
      \kappa_{AB} = -\tfrac{1}{2}n(\obar{g}_{AB})
                  = -\tfrac{1}{2}N^{-1}(\obar{g}_{AB})^{\cdot}
                    -N^{-1}A\nu^1 \lambda_{AB} \\
      \tr\kappa = -\tfrac{2}{r}n(r)
      \komma
   \end{gather*}
   reflecting the definition of the
   area radius $r$ as a volume measure and the second fundamental forms as
   its rate of change. In the above formulas $\obar{g}$ is considered as
   intrinsic to $F$ (and we maintain this from now on) and  
   $\obar{g}^{AB}=r^{-2}\tilde{g}^{AB}$ denotes the inverse of
   $\obar{g}_{AB}$. Differentiating $\tilde{g}$
   directly one gets from its relation to $\obar{g}$ and the formulas above
   the following representations of the trace free parts
   of the second fundamental forms:
   \begin{align*}
      & \tilde{\lambda}_{AB} = -\tfrac{1}{2}r^2\, m(\tilde{g}_{AB}) 
      && \tilde{\lambda}^{AB} = \tfrac{1}{2}r^{-2}\, m(\tilde{g}^{AB}) \\
      & \tilde{\kappa}_{AB} = -\tfrac{1}{2}r^2\, n(\tilde{g}_{AB}) 
      && \tilde{\kappa}^{AB} = \tfrac{1}{2}r^{-2}\, n(\tilde{g}^{AB}) 
   \end{align*}
\item the connection in the normal bundle $T^{\perp}F$. This can be
   represented by a single one form $\eta$, defined as
   \begin{equation}\label{e.eta}
      \begin{aligned}
         \eta(v) & = g(\viernabla_v n,m) = -g(\viernabla_v m,n) \\
                 & = -\frac{1}{2}\,g([n,m],v)
      \end{aligned}
      \qquad v \in TF
   \end{equation}
   From this formula one can see, that $\eta=0$ is equivalent to 
   $[n,m] \in T^{\perp}F$, thus 
   $[T^{\perp}F,T^{\perp}F] \subset T^{\perp}F$ and the theorem of
   Frobenius tells us, that $T^{\perp}F$ is an integrable distribution 
   of two planes in $TM$. If $T^{\perp}F$ is integrable, then we can
   decompose the spacetime into a direct sum 
   $(M,g)=(F^{\perp},\ubar{g})+(F,\obar{g})$, with 
   $g=\ubar{g} \oplus \obar{g}$. Sufficient for the existence of an
   integral manifold $F^{\perp}$ of the distribution $T^{\perp}F$ is
   the Gowdy--type symmetry $y^A \longmapsto -y^A$, $A=2,3$, since then it
   follows $\eta=0$ immediately, because $v \longmapsto \eta(v)$ is an
   antisymmetric map.
\end{enumerate}

Whether or not $T^{\perp}F$ is integrable, we can orthogonally split
the tensor bundle over $M$, following Kundu (\cite{k}).
To perform this task we start with the two 
dimensional Riemannian manifold $(F,\obar{g})$. 
With our choice of spacetime coordinates we have Killing fields
$Y_A=\d_A$. Their spacetime components define projection operators 
\begin{align*}
   & {p_A}^{\mu} := {Y_A}^{\mu} \,, &&
     p_A = \d_A \\
   & {p^A}_{\mu} := \obar{g}^{AB}g_{\mu\nu} {p_B}^{\nu} \,, &&
     p^A = \obar{\nu}^A dt + dy^A
   \komma
\end{align*}
where $\obar{\nu}^A:=\obar{g}^{AB}\,\obar{\nu}_B=\nu^A$, and the
metric components $g_{1B}$ are zero by our definition of the
coordinates.\zeile 
The projection operators into $T^{\perp}F$ can now defined as follows:
Let the unit normal vectors $n$ and $m$ serve as projection operators
${q_a}^{\mu}$ and define their duals by the relations
${q_a}^{\mu} {q^b}_\mu = \delta_a^b$ and
${p_A}^{\mu} {q^b}_{\mu} = 0$. The result is
\begin{align*}
   & q_0 = n = N^{-1}(\d_t-\nu^i\d_i) \,, && q_1 = m = A^{-1}\d_1\\
   & q^0 = N\,dt \,, && q^1 = A \nu^1 dt + A\,dx
\end{align*}
Now we are ready to define the transversal metric $\ubar{g}$ in
$T^{\perp}F$ by 
\begin{equation}\label{e.gubar2}
   \ubar{g}_{ab} = g_{\mu\nu}{q_a}^{\mu}{q_b}^{\nu}
   \komma
\end{equation}
thus in the given frame $\ubar{g}$ and its inverse are represented by the
two dimensional Minkowski metric and from now on we consider $\ubar{g}$ as
an intrinsic object in the tensor bundle over $T^{\perp}F$.\zeile
In summary we have constructed a complete set of projection operators, 
projecting orthogonally tensors over $(TM,g)$ into $(TF,\obar{g})$ and 
$(T^{\perp}F,\ubar{g})$, characterized by the relations
\begin{equation*}
   {p_A}^{\mu}{p^A}_{\nu} + {q_a}^{\mu}{q^a}_{\nu} =
   \delta^{\mu}_{\nu}
   \quad\text{(completeness)}
   \komma
\end{equation*}\vspace{-2.5em}
\begin{align*}
   & {p_A}^{\mu}{p^B}_\mu = \delta_A^B &&
     {p_A}^{\mu}{q^b}_\mu = 0 \\
   & {q_a}^{\mu}{p^B}_{\mu} = 0 &&
     {q_a}^{\mu}{q^b}_{\mu} = \delta_a^B
   \quad\text{(orthogonality)}
   \komma
\end{align*}
with $p_A=\d_A$ and $q_0 \perp q_1$ (these two relations fix the
component representations shown above).

Now we define the
convention already used in some expressions above, that wherever
confusion might arise, we attach a bar to quantities, which will 
be considered as intrinsic to the associated bundle, and a tensor index
furnished with such a bar denotes projected
components, which also can be considered as intrinsic. For
example $\ubar{T}_{ab}$ denotes a tensor in the bundle
$(T^{\perp}F,\ubar{g})$, but 
$T_{\ubar{a}\ubar{b}}={q_a}^{\mu}{q_b}^{\nu}T_{\mu\nu}$, too, although
$T$ might be a spacetime tensor. The philosophy lying beyond this
notation is, that 
quantities with a bar attached to them or to their indices can be
manipulated 
by the associated metric, while indices without a bar always
denote component indices corresponding to the bundle the tensor is
intrinsic to. If we apply this notational convention to the 
metric itself we get the definitions for $\ubar{g}$ and $\obar{g}$ back,
for example we get the identities
$g_{\ubar{a}\ubar{b}}={q_a}^{\mu}{q_b}^{\nu}g_{\mu\nu}=\ubar{g}_{ab}$ and
$g_{\obar{A}\obar{B}}={p_A}^{\mu}{p_B}^{\nu}g_{\mu\nu}=\obar{g}_{AB}$
In particular we have the relations 
\begin{gather*}
   \ubar{\nabla}_a = {q_a}^{\mu}\; \viernabla_{\mu} \\
   \obar{\nabla}_A = {p_A}^{\mu}\; \viernabla_{\mu}
   \komma
\end{gather*}
defining the Levi-Civita connection in the projected bundles, 
and the algebraic identity
\begin{equation*}
   T^{\mu}_{\mu} = T^{\ubar{a}}_{\ubar{a}} + T^{\obar{A}}_{\obar{A}}
   \punkt
\end{equation*}

For later use we need the projected components of the Ricci tensor.
First note, that the normal connection $\eta$ on $(F,\obar{g})$ reads
in our new notation
\begin{equation*}
   \obar{\eta}^A = -\tfrac{1}{2}\,[q_0,q_1]^{\obar{A}}
                 = -\tfrac{1}{2}\,{p^A}_{\mu}[q_0,q_1]^{\mu}
                 = \tfrac{1}{2}\, dp^A(q_0,q_1) 
   \komma
\end{equation*}
or equivalently $\epsilon_{ab}\obar{\eta}^A=1/2\, dp^A(q_a,q_b)$, 
making use of the relation 
$0=\nabla ({p^A}_{\mu}{q_a}^{\mu})$, which gives
$\nabla {q_a}^{\mu} = -{q_a}^{\nu}{p_B}^{\mu}\nabla {p^B}_{\nu}$. 
The $\epsilon_{ab}$ here has its standard meaning 
as the totally antisymmetric symbol (independent of the frame
used, so it is not necessary here to attach a bar to it). On
$T^{\perp}F$ we find 
$dp_A=d(\obar{g}_{AB}\,p^B)=\obar{g}_{AB}\,dp^B$, so that index
manipulations with $\obar{g}$ can be applied, as desired.\zeile
After some calculations we arrive at formulas for the projected
components of the Ricci tensor:
\begin{subequations}
\begin{gather}
   \label{e.Rubar}
   \vierR_{\ubar{a}\ubar{b}} 
     = \ubar{R}_{ab} - 2 r^{-1} \ubar{\nabla}_a\ubar{\nabla}_b\, r
     + 2 r^{-2}(\ubar{\nabla}_a r)(\ubar{\nabla}_b r)
     + \tfrac{1}{4} (\ubar{\nabla}_a\, \obar{g}_{AB})
                    (\ubar{\nabla}_b\, \obar{g}^{AB})
     + 2 \ubar{g}_{ab} \abs{\eta}^2 \\
   \vierR_{\obar{A}\ubar{b}}
     = r^{-2} {\epsilon^c}_b \ubar{\nabla}_c (r^2 \obar{\eta}_A) \\
   \label{e.Robar}
   \vierR_{\obar{A}\obar{B}}
     = -\tfrac{1}{2}\obar{g}_{AD}\ubar{\nabla}^c
        (\obar{g}^{CD}\ubar{\nabla}_c\,\obar{g}_{BC})
     - r^{-1} (\ubar{\nabla}^c r)(\ubar{\nabla}_c\, \obar{g}_{AB})
     - 2\obar{\eta}_A\,\obar{\eta}_B
\end{gather}
\end{subequations}
Contraction of the last equation yields a formula for the Laplacian of
$r$ (Note, that the Laplacian is really a wave operator in the present
situation):
\begin{equation}\label{e.Laplacerubar}
   \ubar{\Delta}r = \tfrac{1}{2}\vierR^{\obar{A}}_{\obar{A}}
      + r^{-1}(\ubar{\nabla}_c r)(\ubar{\nabla}^c r) 
      + r \abs{\obar{\eta}}^2
\end{equation}

Finally we consider other frames of reference and their relation to the
frames already in use in the projected spaces. At first we concentrate on
$(TF,\obar{g})$, which is canonically endowed with the coordinate vector
fields $\d_A$. Alternatively we can introduce the orthonormal frame
$\{\obar{e}_A\}$, defined by $\obar{e}_A := r^{-1}\tilde{e}_A$, with
\begin{alignat*}{4}
    \tilde{e}_2 & = & 
                    e^{-W/2}\cosh V/2 \;\d_2 \quad
                & - & \quad e^{W/2}\sinh V/2 \;\d_3 && \\
    \tilde{e}_3 & = - & 
                    e^{-W/2}\sinh V/2 \;\d_2 \quad
                & + & e^{W/2}\cosh V/2 \;\d_3 &&
   \punkt
\end{alignat*}
The dual frame $\{\obar{\sigma}^A\}$ is defined by 
$\obar{\sigma}^A := r\tilde{\sigma}^A$, with 
$(\tilde{\sigma}^A)_{\mu}(\tilde{e}_B)^{\mu}=\delta^A_B$, thus 
\begin{alignat*}{4}
   \tilde{\sigma}^2 & = &
                   e^{W/2}\cosh V/2 \;dy^2 \quad
               & + & \quad e^{-W/2}\sinh V/2 \;dy^3 && \\
   \tilde{\sigma}^3 & = & e^{W/2}\sinh V/2 \;dy^2 \quad
               & + & e^{-W/2}\cosh V/2 \;dy^3 &&
   \punkt
\end{alignat*}
Finally we introduce the two dimensional quotient manifold
$B:=M/(U(1) \times U(1))$. Then we have isomorphisms
$TB \simeq TM/TF \simeq T^{\perp}F$ and we can consider the metric
$\ubar{g}$ as acting on $TB$, so having constructed the two-dimensional
(quotient) spacetime $(B,\ubar{g})$. If we take $(t,x)$ as coordinates of
$B$, then the metric $\ubar{g}$ has coordinate components
$\ubar{g}_{ab} = -(q^0)^2 + (q^1)^2 
               = (-N^2+(A\nu^1)^2)\,dt^2 + 2A^2\nu^1\,dt\,dx + A^2\,dx^2$, 
giving an alternative way to describe tensor components in 
$(T^{\perp}F,\ubar{g})$, even if this bundle is not integrable. 
The dual base $\{\ubar{e}_a\}$ to $\{\ubar{\sigma}^a=q^a\}$ in
$(B,\ubar{g})$ has the coordinate components 
$\ubar{e}_0=N^{-1}(\d_t-\nu^1\,\d_x)$, 
$\ubar{e}_1=A^{-1}\,\d_x$.
The component notation just 
introduced conflicts with the conventions described above, and in the
following sections we are forced to make clear, which conventions we will
follow.
\subsubsection{The 3+1--geometry}\label{ss.3+1}
To describe the 3--geometry of an arbitrary symmetric Cauchy surface $S$,
we construct the coordinate system $(x',y^A)$ explicitly. On the 
surface of symmetry $F \subset S$ we have already coordinates, such
that the scaled metric $\tilde{g}$ takes the form \eqref{e.gtilde2}.
In a Gaussian neighbourhood of $F$ the metric $h$ of $S$ then has 
the form $h = d{x'}^2 + \sqrt{\abs{h}} \tilde{g}(x')$. Now we introduce a
scale factor 
$a=2\pi \left( \int_0^L \abs{h(z)}^{-1/4}\,dz \right)^{-1}$, where $L$ 
denotes the length of a (projected) geodesic along all of $S^1$ orthogonal
to the orbits $F$. Define a new coordinate $x$ by 
$x(x') = a \int_0^{x'} \abs{h(z)}^{-1/4}\,dz$, then we get a convenient
representation of $h$ as
\begin{equation}\label{e.h2}
   h(x) = A(x)^2 
          \left(
            dx^2 + a^2 \tilde{g}(x)
          \right)
   \komma
\end{equation}
with $A(x)=a^{-1}\abs{h(x)}^{1/4}$ defined on $S^1$, and $\tilde{g}$
transforms under some element of $\Gl(2,\ZZ)$ like $\obar{g}$ in
\eqref{e.gobartransform}. By the way, from this explicit formula and the
definition of $\obar{g}$ in \eqref{e.gobar2} it follows easily that the
relation $r=Aa$ holds.\zeile
The Laplacian of $S$ acting on functions $\psi$ then has the
form
\begin{align*}
   \Delta\psi & = -h^{11}\nabla_1\nabla_1\psi + h^{AB}\Gamma_{AB}^1\psi'
                +\obar{\Delta} \psi\\
   & = -A^{-2} (\psi'' + A^{-1}A'\psi') + \obar{\Delta}\psi
   \komma
\end{align*}
with $\obar{\Delta}\psi=-h^{AB}\d_A\d_B\psi$ since
$\obar{\Gamma}_{AB}^C=0$.

Finally we cast the second fundamental form $k$ of $S$ into a
convenient form. Set $K:=k(m,m)=A^{-2}k_{11}$ and observe 
$k_{1B}=-\viernabla_B n_1=-A\eta_B$ we get
\begin{equation}\label{e.k2}
   k(x) = A^2K\, dx - 2A\eta_B\, dx\,dy^B 
          + \kappa_{AB}\,dy^A\,dy^B
   \komma
\end{equation}
where all quantities on the right-hand side depend on $x$. Taking the
trace yields the relation
\begin{equation*}
   H-K = \tr\kappa
   \punkt
\end{equation*}
\subsubsection{The 4--geometry}\label{ss.4geometry}
We will describe the 4--geometry (locally) in terms of a PMC foliation
by symmetric Cauchy surfaces $(\Sigma,h)$.
It turns out, that there is not much left to do. Of course, we need
the strong energy condition in $M$, and we can choose $\Sigma$ without loss
of generality to have non-vanishing second fundamental form. In view of
theorem 2.2 of part I this 
is enough to guarantee the existence of a local in time PMC foliation of a
neighbourhood of $\Sigma$ in $M$.
The remaining question is, if the leaves of the
foliation turn out to be symmetric. But the  
arguments given in the corresponding place in \cite{h3}
do apply to the present situation, since the Laplacian of $\Sigma$
splits into the Laplacian of $F$ and a part depending only on $x$,
which coincides with the one in the plane symmetric case and we end up 
with
\begin{proposition}\label{prop.4geometry2}\ein
   Let $(M,g)$ be a globally hyperbolic, spatially compact spacetime
   with local $U(1) \times U(1)$ symmetry, obeying the strong
   energy condition.\zeile
   Then there exists a local in time PMC foliation $\{S_t\}$ in $M$,
   covering a neighbourhood $]t_1,t_2[ \times \Sigma$ of $\Sigma=S_0$,
   in $M$. Moreover, if $\Sigma$ is symmetric, then all the leaves of
   the foliation are symmetric, too, and there are coordinates
   $(x^{\mu})=(t,x,y^2,y^3)$ adapted to the foliation, which cast the
   metric into the form
   \begin{equation}
      \label{e.g2}
        g = \begin{pmatrix}
               -N^2+(A\nu^1)^2 + \abs{\obar{\nu}}^2 & A^2\nu^1 & 
               \obar{\nu}_2  & \obar{\nu}_3 \\
               A^2\nu^1        & A^2      & 0             & 0 \\
               \obar{\nu}_2    & 0        & \obar{g}_{22} & \obar{g}_{23} \\
               \obar{\nu}_3    & 0        & \obar{g}_{32} & \obar{g}_{33}
            \end{pmatrix}
      \komma
   \end{equation}
   where
   $N, A, \nu^1$, 
   $\nu^A=\obar{\nu}^A$, $\obar{\nu}_A=\obar{g}_{AB}\,\obar{\nu}^B$, 
   $\obar{g}=r^2\tilde{g}$ are functions on          
   $]t_1,t_2[ \times S^1$, with $r=Aa$ and $a=a(t)$ only. The quantities
   with an overbar transform under some representation of
   $\Gl(2,\ZZ)$ after each transition $x \longmapsto x+2\pi$ in $S^1$. The
   shift functions are fixed by the conditions $\nu^1(t,0)=\nu^1(t,2\pi)=0$
   and $\nu^A(t,0)=0$.
\end{proposition}

The orthonormal frames $\{\viere_{\mu}\}$ and $\{\viersigma^{\mu}\}$
associated with the $3+1$--split read explicitly 
\begin{align*}
   \viere_0 & = n = N^{-1}(\d_t -\nu^i\d_i) \\
   \viere_1 & = m = A^{-1}\,\d_1 \\
   \viere_2 & = e_2 = (Aa)^{-1} \left(+e^{-W/2}\cosh V/2 \;\d_2
                                \;-\;e^{W/2} \sinh V/2 \;\d_3\right) \\
   \viere_3 & = e_3 = (Aa)^{-1} \left(-e^{-W/2}\sinh V/2 \;\d_2
                                \;+\;e^{W/2} \cosh V/2 \;\d_3\right)\\
   \viersigma^0 & = N\,dt \\
   \viersigma^1 & = A(\nu^1\,dt + dx)\\
   \viersigma^2 & = (Aa) 
      \left(e^{W/2}\cosh V/2\, (\obar{\nu}^2\,dt+dy^2) 
      \;+\; e^{-W/2}\sinh V/2\, (\obar{\nu}^3\,dt+dy^3)
      \right) \\
   \viersigma^3 &= (Aa) 
      \left(e^{W/2}\sinh V/2\, (\obar{\nu}^2\,dt+dy^2) 
      \;+\; e^{-W/2}\cosh V/2\, (\obar{\nu}^3\,dt+dy^3)
      \right)
\end{align*}
Finally, the Ricci rotation coefficients will be calculated. Some 
expressions are not explicitly given, since they turn out to be too
complicated. Instead, the dependence on the geometric quantities involved
will be specified in square brackets: 
{\samepage
\begin{alignat*}{2}
  & \viergamma_{01}^0\: &&=\: \viergamma_{00}^1 = A^{-1}N^{-1}N' \\
  & \viergamma_{0B}^0 &&= \viergamma_{00}^B = 0 \\
  & \viergamma_{11}^0 &&= \viergamma_{01}^1 = -K \\
  & \viergamma_{1B}^0 &&= \viergamma_{10}^B 
                        = (Aa)^{-1} {\obar{e}_B}^C\eta_C \\
  & \viergamma_{AB}^0 &&= \viergamma_{A0}^B 
                        = -{\obar{e}_A}^C{\obar{e}_B}^D\,\kappa_{CD}\\
  & \viergamma_{01}^1 &&= 0\\
  & \viergamma_{0B}^1 &&= \viergamma_{0B}^1
      [N^{-1},\obar{\nu},\obar{\nu}',A,A^{-1},A',a,a^{-1},
       V,W,V',W',\dot{V},\dot{W}]
      = -\viergamma_{01}^B \\
  & \viergamma_{0B}^C &&= \viergamma_{0B}^C
      [N^{-1},\nu^1,A,A^{-1},A',a,a^{-1},V,W,V',W',\dot{V},\dot{W}] \\
  & \viergamma_{ij}^k &&= \gamma_{ij}^k[A^{-1},A',V,W,V',W']
   \komma
\end{alignat*}
where the dependence on the derivatives of $V$ and $W$ is linear.
}
%
%
%
%
%
%
%
%
%
%
\subsection{The f\/ield equations}
Again we represent the matter quantities by $\rho=T(n,n)$,
$j=-T(n,m)=A^{-1}j_1$ and $S_{ij}=T(\viere_i,\viere_j)$. 
\subsubsection{2+1--decomposition of the constraints}
First consider the Hamiltonian constraint 
\begin{equation*}
   R+H^2-\abs{k}^2=16\pi\rho
   \punkt
\end{equation*}
Decomposing $R$ by the Gauss--Codazzi formula and $k$ by \eqref{e.k2} we
get the following expression:
\begin{equation}\label{e.ch2+2}
   m(\tr\lambda) = 8\pi\rho - H\tr\kappa +
       \abs{\eta}^2 
       + \tfrac{1}{2}
           \bigl(\tfrac{3}{2}(\tr\lambda)^2+\abs{\tilde{\lambda}}^2\bigr)
       + \tfrac{1}{2}
           \bigl(\tfrac{3}{2}(\tr\kappa)^2+\abs{\tilde{\kappa}}^2\bigr)
   \punkt
\end{equation}
The first component of the momentum constraint
\begin{equation*}
   \nabla^i k_{i1} - \nabla_1 H = 8\pi j_1
\end{equation*}
can be decomposed into 
\begin{equation}\label{e.cm2+11}
   m(\tr\kappa) = -8\pi j_1 - H \tr\lambda 
                  +\tfrac{3}{2}\tr\kappa\tr\lambda
                  +\tilde{\lambda}^{AB}\tilde{\kappa}_{AB}
\end{equation}
(where the symmetry forced $\div \eta=0$).
The other components of the momentum constraints are calculated
analogously to (again some terms cancel out due to symmetry)
\begin{equation}\label{e.cm2+1B}
   m(\eta_B) = -8\pi j_B + (\tr\lambda)\eta_B
   \punkt
\end{equation}

As in \cite{h3} a convenient form of the equation is
achieved by introducing the null expansions
\begin{equation}\label{e.expansions2}
   \vartheta_{\pm} := - ( \tr\lambda \pm \tr\kappa)
                    = \frac{2}{r} k_{\pm}(r)
\end{equation}
(compare the formulas for $\lambda$ and $\kappa$ in \ref{ss.2+2}), which
yields an easy formula for the Hawking mass
\begin{equation}\label{e.mass2}
   \mass := m_H(F) = -\tfrac{1}{8}r(r\vartheta_+)(r\vartheta_-)
          = -\tfrac{1}{2}r(\viernabla^{\alpha}r)(\viernabla_{\alpha}r)
   \komma
\end{equation}
by 
$\vartheta_+\vartheta_-=4/r^2(\viernabla^{\alpha}r)(\viernabla_{\alpha}r)$.

Adding and subtracting the constraint equations now yields
\begin{equation}\label{e.mvartheta2}
   m(\vartheta_{\pm}) = 
      - 8\pi(\rho \mp j_1) 
      - \abs{\eta}^2
      - \tfrac{1}{2}\abs{\tilde{\lambda} \pm \tilde{\kappa}}^2
      \mp H\vartheta_{\pm} - \tfrac{3}{4}\vartheta_{\pm}^2
\end{equation}
and the transition to the variables $\omega_{\pm} := r\vartheta_{\pm}$ casts
this equation into
\begin{equation}\label{e.momega2}\begin{split}
   m(\omega_{\pm}) & = m(r)\vartheta_{\pm} + r m(\vartheta_{\pm})\\
     & = -8\pi r(\rho\mp j_1) - r\abs{\eta}^2  
         -\tfrac{1}{2}r\abs{\tilde{\lambda} \pm \tilde{\kappa}}^2
         \mp H\omega_{\pm}
         +\tfrac{1}{4\pi}
         \left( \omega_+ \omega_- - 2\omega_{\pm}^2 \right)
   \komma
\end{split}\end{equation}
where the identity
$m(r)\vartheta_{\pm}=\tfrac{1}{4\pi}((\omega_+ +\omega_-)\omega_{\pm})$
has been used.
\subsubsection{2+2--decomposition of the f\/ield equations}
\label{ss.field2+2}
With the 2+2--geometry in mind, where we denote tensor components with
respect to the frame of projection operators $\{q_a,p_A\}$, we
write the  
equation \eqref{e.Rubar} for $\vierR_{\ubar{a}\ubar{b}}$ in terms of 
$\tilde{g}$,
\begin{equation*}
   \vierR_{\ubar{a}\ubar{b}} 
     = \ubar{R}_{ab} - 2 r^{-1} \ubar{\nabla}_a\ubar{\nabla}_b\, r
     + \tfrac{1}{4} (\ubar{\nabla}_a\, \tilde{g}_{AB})
                    (\ubar{\nabla}_b\, \tilde{g}^{AB})
     + 2 \ubar{g}_{ab} \abs{\eta}^2
   \komma
\end{equation*}
as a field equation in the quotient spacetime $(B,\ubar{g})$ for the radius
function $r$. We still have to 
eliminate the unknown $\ubar{R}_{ab}=-\ubar{K}\ubar{g}_{ab}$, where
$\ubar{K}:=\ubar{R}_{0101}$ denotes the Gaussian curvature of
$(B,\ubar{g})$ in $(M,g)$. Contracting the
equation for $\ubar{R}_{ab}$ and replacing $\ubar{\Delta}r$ by
\eqref{e.Laplacerubar} yields an equation for $\ubar{K}$: 
\begin{equation*}
   \ubar{K}
     = \tfrac{1}{2}
       (-\vierR^{\ubar{c}}_{\ubar{c}} + \vierR^{\ubar{A}}_{\ubar{A}})
       + r^{-2}(\ubar{\nabla}^c r)(\ubar{\nabla_c r})
       + \tfrac{1}{8}
       (\ubar{\nabla}^c\tilde{g}_{AB})(\ubar{\nabla}_c\tilde{g}^{AB})
       +3\abs{\eta}^2
\end{equation*}
Inserting this into the equation for $\vierR_{\ubar{a}\ubar{b}}$ results in
\begin{equation*}\begin{split}
   \ubar{\nabla}_a\ubar{\nabla}_b r = &
      -\tfrac{1}{2}r^{-1}
         (\ubar{\nabla}^c r)(\ubar{\nabla}_c r)\ubar{g}_{ab}
      -\tfrac{1}{2}r\Bigl(
         \vierR_{\ubar{a}\ubar{b}} -\tfrac{1}{2}
            (\vierR^{\ubar{c}}_{\ubar{c}}
             -\vierR^{\obar{A}}_{\obar{A}})\ubar{g}_{ab}
       \Bigr)\\ & 
      +\tfrac{1}{8}r\Bigl(
          (\ubar{\nabla}_a\tilde{g}_{AB})(\ubar{\nabla}_b\tilde{g}^{AB})
          -\tfrac{1}{2}
          (\ubar{\nabla}^c\tilde{g}_{AB})(\ubar{\nabla}_c\tilde{g}^{AB})
             \ubar{g}_{ab}
       \Bigr)
      -\tfrac{1}{2}r\abs{\eta}^2\ubar{g}_{ab}
\end{split}\end{equation*}
Now we are going to interpret the first three terms appearing on the
right-hand side of the field equation: Recalling the definition of the mass
function $\mass$ in \eqref{e.mass2} we can write the first term as 
$\tfrac{\mass}{r^2}\ubar{g}_{ab}$. The curvature expression in the second
term can be rewritten using Einstein's equations in terms of the projected
energy momentum tensor $\ubar{T}$ in the form 
$8\pi(\ubar{T}_{ab}-\tr\ubar{T}\ubar{g}_{ab})$.\zeile
To simplify the third term, there is more structure involved than already
presented. The space of two-dimensional metrics $\text{Bil}(2,\RR)$ is a
three dimensional real 
vector space, whose elements we denote by
$\tilde{\phi}=\tilde{\phi}^i\,E_i$, where the 
basis $\{E_i\}$ has been chosen as 
$E_1=\left(\begin{smallmatrix}1 & 0 \\ 0 & 1\end{smallmatrix}\right)$,
$E_2=\left(\begin{smallmatrix}1 & 0 \\ 0 & -1\end{smallmatrix}\right)$,
$E_3=\left(\begin{smallmatrix}0 & 1 \\ 1 & 0\end{smallmatrix}\right)$.
In this picture the determinant is a quadratic form on $\text{Bil}(2,\RR)$,
and the matrix representation with respect to the given basis turns out to be
$-\eta_{ij}$, where $\eta_{ij}=\text{diag}(-++)$ denotes the standard
Minkowski metric in three dimensional Minkowski space. The restriction of
this space to the hypersurface 
$\eta(\tilde{\phi},\tilde{\phi})=-1 \Longleftrightarrow \det\tilde{\phi}=1$ 
is the well known hyperbolic plane $H^2$, and we have established a
correspondence between the two-dimensional unit-determinant metrics
$\tilde{g}$ and the elements of the hyperbolic plane. 
Considering $\tilde{\phi}$ as a function on the quotient manifold $B$,
introduced in \ref{ss.2+2} we have a map
$\tilde{\phi}:(B,\ubar{g}) \longrightarrow (H^2 \subset \RR^{2,1},\eta_{ij})$, 
$(t,x) \longmapsto \tilde{\phi}^i(\tilde{g}(t,x))E_i$.
Now let us interpret $\tilde{\phi}$ for a moment as a wave map from $B$ to
$H^2$, then the energy momentum tensor $\leftsuper{\phi}\ubar{T}$
associated with $\tilde{\phi}$ reads
\begin{equation}\label{e.Tphi}
   \leftsuper{\phi}\ubar{T}_{ab}
     = \ubar{\nabla}_a \tilde{\phi}^i \ubar{\nabla}_b \tilde{\phi}_i
       -\tfrac{1}{2}
       (\ubar{\nabla}^c \tilde{\phi}^i \ubar{\nabla}_c \tilde{\phi}_i)
       \ubar{g}_{ab}
\end{equation}
and with the correspondences
$\tilde{\phi}_i = \eta_{ij}\tilde{\phi}^j = -\tilde{g}^{AB}$ and 
$\tilde{\phi}_i\tilde{\phi}^i=-\tfrac{1}{2}\,\tilde{g}_{AB}\,\tilde{g}^{AB}$,
we see 
that the third term in the field equation for $r$ can be rewritten as
$-\tfrac{1}{4}r \leftsuper{\phi}\ubar{T}_{ab}$.

Collecting all these things together we get the field equation for $r$ in a
convenient form:
\begin{equation}
   \label{e.fieldr2}
   \ubar{\nabla}_a\ubar{\nabla}_b r 
     =   \frac{\mass}{r^2}\ubar{g}_{ab}
       - 4\pi r (\ubar{T}_{ab}-\tr\ubar{T}\ubar{g}_{ab})
       - \tfrac{1}{4}r\,\leftsuper{\phi}\ubar{T}_{ab}
       - \tfrac{1}{2}r\abs{\eta}^2\ubar{g}_{ab}
   \punkt
\end{equation}
Differentiating the mass function we find for the mass flux equation the
formula 
\begin{equation}
   \label{e.massflux2}
   \ubar{\nabla}_a \mass = 
   \Bigl(
     4\pi r^2 (\ubar{T}_{ab}-\tr\ubar{T}\ubar{g}_{ab})
     +\tfrac{1}{4}r\,\leftsuper{\phi}\ubar{T}_{ab}
   \Bigr)(\ubar{\nabla}^b r)
   + \tfrac{1}{2}r^2\abs{\eta}^2\ubar{\nabla}_a r
   \punkt
\end{equation}

The contribution of $\leftsuper{\phi}\ubar{T}$ to the right-hand sides of
these equations is known in so far, that $\leftsuper{\phi}\ubar{T}$ obeys
the dominant and strong energy 
conditions. Furthermore, we can again rewrite its components using the
formulas for the trace free parts of $\lambda$ 
and $\kappa$ in \ref{ss.2+2} 
(remember, that $\ubar{\nabla}_0$, $\ubar{\nabla}_1$ act as $n$,
respectively $m$, on $\tilde{g}$, in the formula for
$\leftsuper{\phi}\ubar{T}_{ab}$). The result is 
\begin{subequations}
\begin{gather}
   \leftsuper{\phi}\rho = \leftsuper{\phi}p 
     = \abs{\tilde{\lambda}}^2 + \abs{\tilde{\kappa}}^2 
   \label{e.rhophi} \\
   \leftsuper{\phi}j_{\ubar{1}} 
     = -2\tilde{\lambda}_{AB}\tilde{\kappa}^{AB}
   \label{e.jphi}
\end{gather}
\end{subequations}

Now we want to analyse the underlying wave map. To make things more
explicit, consider the parameters $V,W$ of $\tilde{g}$ as coordinates
of $H^2$ in view of the correspondence $\tilde{\phi}^iE_i$. The pullback
under this parametrization yields a representation $\hat{h}$ of the metric
$(\eta_{ij})_{|H^2}$, with the explicit form 
$\hat{h}_{IJ} = dV^2 + (\cosh V)^2\,dW^2$, leading to 
$\hat{\Gamma}_{23}^3=\tanh V$ and 
$\hat{\Gamma}_{33}^2=-\sinh V \cosh V$
as non-vanishing Christoffel symbols. The pullback of the map 
$\tilde{\phi}$ under this parametrization induces a map
$\phi:(B,\ubar{g}) \longrightarrow (H^2,\hat{h})$, 
$\phi(t,x)=(V(t,x),W(t,x))$. 
The wave map $\phi$ is then defined to obey
\begin{equation}\label{e.wavemap}
   \ubar{\nabla}^c \ubar{\nabla}_c \phi^K
   + \hat{\Gamma}_{IJ}^K \ubar{\nabla}^c \phi^I \ubar{\nabla}_c \phi^J
   = 0
   \punkt
\end{equation}

Let us turn back to the field equations. Writing equation \eqref{e.Robar}
in the form
\begin{equation*}
   \vierR_{\obar{A}\obar{B}} + 2\obar{\eta}_A\obar{\eta}_B
     = -\tfrac{1}{2} r^{-2}\obar{g}_{AD} 
          \ubar{\nabla}^c (r^2\obar{g}^{CD}\ubar{\nabla}_c\obar{g}_{BC})
     = -\tfrac{1}{2} \tilde{g}_{AD} 
          \ubar{\nabla}^c (\tilde{g}^{CD}\ubar{\nabla}_c r^2\tilde{g}_{BC})
\end{equation*}
we arrive after some calculation at
\begin{equation*}\begin{split}
   \ubar{\nabla}^c(r^2 \ubar{\nabla}_c \tilde{g}_{AB}) = &
     r^2 \tilde{g}^{CD} 
       (\ubar{\nabla}^c \tilde{g}_{AD})(\ubar{\nabla}_c \tilde{g}_{BC})\\
   & -16\pi \big(\obar{T}_{AB} 
                 -\tfrac{1}{2}\tr\obar{T}\obar{g}_{AB}\big)
     - 4 \big(\obar{\eta}_A\obar{\eta}_B 
              -\tfrac{1}{2}\abs{\eta}^2\obar{g}_{AB}\big)
   \komma
\end{split}\end{equation*}
where $\obar{T}$ denotes the orthogonal projection of the energy momentum
tensor $T$ into $(F,\obar{g})$. More explicitly we get
\begin{subequations}\label{e.fieldVWa}
\begin{gather}
\begin{split}
   \ubar{\nabla}^c ( r^2 \ubar{\nabla}_c V) = &
     r^2 \sinh V \cosh V (\ubar{\nabla}^c W)(\ubar{\nabla}_c W)\\
  &  -2r^2 (\cosh V)^{-1}
     \Big(
       (T_{23} - \tfrac{1}{2}\tilde{g}^{AB}T_{AB}\tilde{g}_{23})
       -\tfrac{1}{2}
         (\eta_2\eta_3 - \tfrac{1}{2}\tilde{g}^{AB}\eta_A\eta_B\tilde{g}_{23})
     \Big)
\end{split}\\
\begin{split}
   \ubar{\nabla}^c ( r^2 \ubar{\nabla}_c W) = &
     -r^2 \tanh V (\ubar{\nabla}^c W)(\ubar{\nabla}_c V)\\
  &  -r^2 (\cosh V)^{-1}
     \Big(
       (e^{-W} T_{22} - e^W T_{33})
       -\tfrac{1}{2}
         (e^{-W} (\eta_2)^2 - e^W (\eta_3)^2)
     \Big)
\end{split}
\end{gather}
\end{subequations}
or 
\begin{subequations}\label{e.fieldVWb}
\begin{gather}
\begin{split}
   \ubar{\nabla}^c \ubar{\nabla}_c V 
   & -\sinh V \cosh V (\ubar{\nabla}^c W)(\ubar{\nabla}_c W) 
     = -2/r (\ubar{\nabla}^c r)(\ubar{\nabla}_c V) \\
   & -2 (\cosh V)^{-1}
     \Big(
       (T_{23} - \tfrac{1}{2}\tilde{g}^{AB}T_{AB}\tilde{g}_{23})
       -\tfrac{1}{2}
         (\eta_2\eta_3 - \tfrac{1}{2}\tilde{g}^{AB}\eta_A\eta_B\tilde{g}_{23})
     \Big)
\end{split}\\
\begin{split}
   \ubar{\nabla}^c \ubar{\nabla}_c W 
    & + \tanh V (\ubar{\nabla}^c W)(\ubar{\nabla}_c V)
      = -2/r (\ubar{\nabla}^c r)(\ubar{\nabla}_c W)\\
    &   - (\cosh V)^{-1}
       \Big(
         (e^{-W} T_{22} - e^W T_{33})
         -\tfrac{1}{2}
           (e^{-W} (\eta_2)^2 - e^W (\eta_3)^2)
       \Big)
\end{split}
\end{gather}
\end{subequations}
The left-hand sides of equations \eqref{e.fieldVWb} coincide with the the
left-hand side of \eqref{e.wavemap} and we conclude, that $V$ and $W$ solve
an inhomogeneous wave map.
\subsubsection{3+1--decomposition of the f\/ield equations}
Now we want to calculate the ADM equations and bring them into a form
most similar to the equations in the plane symmetric case of part I.
The formulas concerning the 3--geometry in \ref{ss.3+1}
are of particular importance for the calculations here, as well as the
definitions of $\lambda$, $\kappa$ and $\eta$ in \ref{ss.2+2}. 
A straightforward calculation then yields the following set of equations.

The constraint equations
\begin{gather}
   \label{e.ch2}
   (A^{1/2})'' 
     = \tfrac{1}{8}A^{5/2}\Big(
         H^2 - \tfrac{1}{2}(H-K)^2 - K^2 
         -\abs{\tilde{\kappa}}^2 -\abs{\tilde{\lambda}}^2 - 2\abs{\eta}^2
         -16\pi\rho
       \Big) \\
   \label{e.cm21}
   K' = -3 A^{-1}A'K + A^{-1}A'H + H' 
        - A\tilde{\kappa}_{AB}\tilde{\lambda}^{AB}  + 8\pi Aj \\
   \label{e.cm22}
   \eta_B^{\prime} = -2A^{-1}A'\eta_B - 8\pi A j_B 
   \punkt
\end{gather}
The equations fixing the foliation
\begin{gather}
   \label{e.l2}
   N'' = -A^{-1}A'N' + A^2N 
      \Big( \tfrac{1}{2}(H-K)^2 + K^2
             +\abs{\tilde{\kappa}}^2 + 2\abs{\eta}^2
             + 4\pi(\rho+\tr S) 
      \Big) -A^2  \\
   \label{e.p2}
   \dot{H} = 1 + \nu^1 H' \\
   \label{e.p'2}
   (\d_t - \nu^1 \d_x)H' = {\nu^1}' H'
   \punkt
\end{gather}
The evolution equations
\begin{gather}
   \label{e.e12}
   \dot{A} = -NAK + A{\nu^1}' + A'\nu^1 \\
   \label{e.e22}
   \dot{a} = -\tfrac{1}{2}Na(H-3K) - a{\nu^1}' \\
   \label{e.e32}
   \begin{split}
      \dot{K} = & \nu^1 K' - A^{-2}(N'' - A^{-1}A'N') \\
         & + N\Big( 
                         -4 A^{-5/2} (A^{1/2})'' + (A^{-2}A')^2 + HK\\
         &\qquad\quad\!  -\abs{\tilde{\lambda}}^2 + 2\abs{\eta}^2
                         -8\pi(A^{-2}S_{11}+\tfrac{1}{2}(\rho-\tr S)) 
      \Big)
   \end{split}\\
   \label{e.e42}
   \begin{split}
      \dot{\eta}_B = &
         \big(\tfrac{1}{2}N(H+K)-2A^{-1}A'\nu^1\big)\eta_B 
          -2N\tilde{\kappa}_{BC}\eta^C \\
       & +\tfrac{1}{2}A^{-2}A'(H-K)\nu_B 
         -\big( \tilde{\kappa}_B^D\tilde{\lambda}_{DC}
               +\tfrac{1}{2}(H-K)\tilde{\lambda}_{BC}
               -A^{-2}A'\tilde{\kappa}_{BC}
          \big)\nu^C\\
       & -8\pi A\nu^1 j_B
   \end{split}
   \punkt
\end{gather}
As in \cite{h3} integration of \eqref{e.e22} over
the circle yields the analogous 
equation for the first component of the shift vector,
\begin{equation}\label{e.s2}
   {\nu^1}' = -\tfrac{1}{2}N(H-3K) + \tfrac{1}{2}\int_{S^1}N(H-3K)
\end{equation}
and the definition of the second fundamental form provides some additional
equations, namely
\begin{equation}\label{e.sB2}
   {\nu^B}' = -2NA\eta^B
\end{equation}
and the coordinate representations of $\lambda$ and $\kappa$ in
\ref{ss.2+2}. The remaining equations for $V$ and $W$ are still
missing. To this end we supplement the system \eqref{e.ch2}--\eqref{e.sB2}
by the field equations \eqref{e.fieldVWa} or \eqref{e.fieldVWb}, and end up
with the full system of equations.

As in the plane symmetric case, we can formulate an existence and
uniqueness theorem for solutions of the equations
\eqref{e.ch2}--\eqref{e.sB2}, \eqref{e.fieldVWb}. First we define a 
{\em symmetric initial data set} for a spacetime with local 
$U(1) \times U(1)$ symmetry by the smooth collection $(\Sigma,h,k)$, where
$\Sigma$ denotes a (possibly non-trivial) torus bundle over the circle
and the fundamental forms 
$h$ and $k$ are represented in a suitable coordinate system as \eqref{e.h2}
and \eqref{e.k2}, respectively. If there are matter fields, then we assume
the matter data and equations to be smooth and symmetric, leading to a well
posed Cauchy problem coupled to the reduced field equations in harmonic
coordinates.\zeile
On the universal cover $\hat{\Sigma}$ of $\Sigma$ the induced data then is
invariant under the action of $G$, hence $G$ acts isometrically on the
whole Cauchy development, which induces a local $U(1) \times U(1)$ symmetry
on the Cauchy development of $(\Sigma,h,k)$. Assuming 
$\lambda:=\abs{k}^2+4\pi(\rho+\tr S)>0$ somewhere on $\Sigma$, we get for
$t_0 \in \RR$ a 
unique symmetric local in time PMC foliation, defined on some time interval 
containing $t_0$, with $\Sigma=S_{t_0}$, and coordinates described in 
proposition \ref{prop.4geometry2} and we just have proved:
\begin{proposition}\label{prop.adm2}\ein
   Let $(\Sigma,h,k)$ be a symmetric initial data set for a spacetime with
   local $U(1) \times U(1)$ symmetry, with matter obeying
   the strong energy condition and $\lambda>0$ somewhere on
   $\Sigma$. Further, let $t_0$ denote an arbitrary real number.\zeile
   Then there exists a $\delta>0$ and a 
   PMC foliated surface symmetric spacetime $(\bar{M},\bar{g})$
   diffeomorphic to  
   $]t_0-\delta,t_0+\delta[\,\times \Sigma$ with an embedding 
   $\iota : \Sigma \longrightarrow \bar{M}$, satisfying
   $\iota(\Sigma)=S_{t_0}$ 
   and $\iota_*h$, $\iota_*k$ are the first and second fundamental form of
   $S_{t_0}$ in $(\bar{M},\bar{g})$. $(\bar{M},\bar{g})$ obeys the strong
   energy condition and $\bar{g}$ can be written in  the form described in
   proposition \ref{prop.4geometry2}. This construction is unique up to
   the choice of $t_0$ and $\delta$.
\end{proposition}
\subsubsection{The expanding model}\label{ss.expanding2}
Now we can proceed in close analogy to the corresponding analysis in
part I. The formula \eqref{e.mass2} for the Hawking mass shows,
that $\grad r$ is timelike as long as $\mass>0$. Indeed, proposition
3.1 in \cite{r} proves this, provided the spacetime is not flat and the
dominant energy condition holds. 

Therefore, under these conditions $\grad r$ is timelike and $\vartheta_+$,
$\vartheta_-$ have fixed and opposite signs. Without loss of
generality we choose the time orientation, such that $\grad r$ is past
pointing and $dr$ is future pointing (by the induced time orientation of
the cotangent bundle). Then $\vartheta_+>0$ and
$\vartheta_-<0$ which classifies the spacetime as expanding in the sense
described in part I.\zeile
Again we expect the singularity in the distant past from our symmetric
initial data Cauchy surface $\Sigma$ and any
symmetric Cauchy surface $S$ in $M$ is not maximal with mean curvature not
everywhere positive on $S$.
%
%
%
%
%
%
%
%
%
%
\subsection{A priori estimates for the f\/ield equations}\label{s.estimates2}
Assume the dominant and strong energy condition to be fulfilled in
$(M,g)$. Let $\Sigma$ be a symmetric Cauchy surface in $M$ with strictly
negative mean curvature $H$ and denote by $\{S_t\}$, $t \in \,]t_1,t_2[$
the local in time PMC  foliation with $\Sigma=S_0$ and the time
orientation chosen in  correspondence with \ref{ss.expanding2}, such that
$H$ decreases with decreasing PMC time.\zeile
We consider the past $D^-(\Sigma)$ of $\Sigma$. In $D^-(\Sigma)$ the mean
curvature is bounded from above by some $\bar{H}<0$, and $H=\bar{H}$ only
on $\Sigma$. Thus $\abs{H}$ is bounded from below and as long as $H$
remains finite, we find the following estimates. 

Consider first the constraint equation in the form
\eqref{e.momega2}. The dominant energy condition gives the
inequality 
\begin{equation*}
      m(\omega_{\pm}) \le
         \mp H\omega_{\pm}
         +\tfrac{1}{4\pi}
         \left( \omega_+ \omega_- - 2\omega_{\pm}^2 \right)
      \komma
\end{equation*}
from which we infer the basic estimate
\begin{equation}\label{e.mom2}
   \abs{\vartheta_{\pm}} \le 4 \abs{H}
   \qquad\Longrightarrow\qquad
   \abs{A^{-2}A'} \le C \,,\quad \abs{K} \le C
   \punkt
\end{equation}
Now we can perform nearly the same estimates, as we have done in the
corresponding place in \cite{h3}. The additional terms
appearing here in the more general equations
\eqref{e.ch2}--\eqref{e.cm21}, \eqref{e.l2}--\eqref{e.e22} and
\eqref{e.s2} do not cause any problems. There is only 
one additional estimate, which will become important in our further
analysis, resulting from the integration of equation \eqref{e.ch2}
along $S^1$, bounding not only 
$\int_{S^1} \rho$, but also 
$\int_{S^1} (\abs{\tilde{\lambda}}^2 + \abs{\tilde{\kappa}}^2
          + \abs{\eta}^2)
  = \int_{S^1} (\leftsuper{\phi}\rho + \abs{\eta}^2)$,
and we can state the analogous  
\begin{proposition}\label{prop.first_estimates2}\ein
   Let $(M,g)$ be a globally hyperbolic, spatially compact spacetime
   with local $U(1) \times U(1)$ symmetry, obeying the dominant and
   strong energy condition. Assume the existence of a symmetric Cauchy
   surface $\Sigma$ with strictly negative mean curvature. In
   particular we get from proposition \ref{prop.4geometry2} a PMC time 
   coordinate $t$, ranging in $]t_1,t_2[$ with $\Sigma=\{t=0\}$ and $H$
   decreases with decreasing $t$.\zeile
   Then we have uniformly on $]t_1,0]$ 
   \begin{equation*}
      \abs{A},\abs{A^{-1}},\abs{A'},\abs{a},\abs{a^{-1}},
      \abs{H},\abs{H'},\abs{K},
      \abs{N},\abs{N^{-1}},\abs{N'},\abs{\nu^1},\abs{{\nu^1}'}
      \le C
      \punkt
   \end{equation*}
\end{proposition}

Until now the control over the coefficients of the fundamental
forms $h$ and $k$ is not complete. We still need bounds for the metric
coefficients $V$, $W$, the components $\eta$ and $\kappa$ of $k$ and the
remaining components $\nu^B$ of the shift vector. We will get bounds for
most of these quantities here in this section, using the bound for
$\int_{S^1} (\rho + \leftsuper{\phi}\rho + \abs{\eta}^2)$.
\begin{proposition}\label{prop.second_estimates}\ein
   Under the hypotheses of proposition \ref{prop.first_estimates2} we get
   uniform bounds for
   \begin{equation*}
      \abs{V},\abs{W}, \abs{\eta_B}, \abs{\nu^B}, \abs{{\nu^B}'}
   \end{equation*}
   on $]t_1,0]$.
\end{proposition}
\begin{proof}\zeile
   The bounds for $\eta_B$, $\nu^B$ and ${\nu^B}'$ are simple consequences
   of the bounds for $V$ and $W$, since having bounded them, we can
   conclude as follows: The bounds for $V$ and $W$ allow coordinate
   components to be bounded in terms of components according to the
   orthonormal frame vectors $\obar{e}_2,\obar{e}_3$ in $(F,\obar{g})$ 
   defined in
   \ref{ss.2+2}. We adopt the convention, that a hat above indices denotes
   components with respect to an orthonormal frame. Having this in 
   mind we see, that $\int_{S^1}\rho$ bounds 
   $\int_{S^1}j_B \le C \int_{S^1} j_{\hat{B}}$ by the dominant energy
   condition, so
   integration of the constraint \eqref{e.cm22} yields a bound for the
   difference $A^2\eta_B\rvert^{x_2}_{x_1}$, which is independent of
   $t$. Using $\eta_B \le C \eta_{\hat{B}} \le C (1+\abs{\eta}^2)$ we can
   bound the integral $\int_{S^1}\abs{A^2\eta_B}$ by 
   $C \int_{S^1}(1+\abs{\eta}^2)$, which together with the estimated
   difference yields the desired estimate for
   $\eta_B$. Now we get immediately a bound for ${\nu^B}'$ by inspection of
   equation \eqref{e.sB2}, and the condition $\nu^B(t,0)=0$ bounds
   $\nu^B$.

   Let us now investigate the field equations \eqref{e.fieldVWa} for $V$
   and $W$, considered as equations on $(B,\ubar{g})$ endowed with the
   coordinates $(t,x)$, as described in the end of
   \ref{ss.2+2}. Then we calculate explicitly in these coordinates
   \begin{equation*}\begin{split}
      \abs{\tilde{\lambda}}^2 + \abs{\tilde{\kappa}}^2
      & = \tfrac{1}{4} ( m(\tilde{g}_{AB})m(\tilde{g}^{AB})
                        +n(\tilde{g}_{AB})n(\tilde{g}^{AB}) ) \\
      & = \tfrac{1}{2}A^{-2} \left( {V'}^2 + (\cosh V)^2 {W'}^2 \right) \\
      & + \tfrac{1}{2}N^{-2} \left( 
             (\dot{V}-\nu^1 V')^2 + (\cosh V)^2 (\dot{W}-\nu^1 W')^2
                             \right)
      \komma
   \end{split}\end{equation*}
   and we see, that the first term on the right-hand side of equations
   \eqref{e.fieldVWa} can be bounded in terms of this expression.\zeile
   For the second term on the right-hand side of \eqref{e.fieldVWa} we
   exploit the special structure appearing there. The first aim is, to
   express the tensor components with respect to an orthonormal
   frame. Looking at the definitions of the coframe $\{\obar{\sigma}^B\}$
   defined in \ref{ss.2+2} we recognize the relations
   \begin{align*}
      e^{W/2} dy^2  & = r^{-1} \left( + \cosh V/2\,\obar{\sigma}^2
                                - \sinh V/2\,\obar{\sigma}^3 \right)\\
      e^{-W/2} dy^3 & = r^{-1} \left( - \sinh V/2\,\obar{\sigma}^2
                                + \cosh V/2\,\obar{\sigma}^3 \right)
      \punkt
   \end{align*}
   If we mark tensor components expressed in this basis by a tilde, we see
   that the terms in brackets in both equations \eqref{e.fieldVWa} are
   easily rewritten with respect to this basis, by simply putting a tilde
   above each tensor component and deleting the factors 
   $e^{\pm W/2}$. But we have schematically
   $\eta = \eta_B dy^B = \tilde{\eta}_B\, e^{\pm W/2}\,dy^B
         = f_B(\cosh V/2) \tilde{\eta}_B\,\obar{\sigma}^B$, with
   $f_B(\cosh V/2) \tilde{\eta}_B=\eta_{\hat{B}}$ is the component
   according to the orthonormal frame. A similar relation holds for
   $T_{AB}$. Since we have $\eta_{\hat{B}} \le 1+\abs{\eta}^2$ and
   $T_{\hat{A}\hat{B}} \le \rho$ we conclude, that the terms in brackets in
   \eqref{e.fieldVWa} are bounded by $C \cosh V (1+\rho+\abs{\eta}^2)$,
   which leads together with our result about the first term on the
   right-hand side of \eqref{e.fieldVWa} to
   \begin{equation*}
      \abs{\ubar{\nabla}^c ( r^2 \ubar{\nabla}_c \phi)}
      \le C (1+\rho+\abs{\tilde{\lambda}}^2+\abs{\tilde{\kappa}}^2
              +\abs{\eta}^2)
       \komma
   \end{equation*}
   and since the integration of the right-hand side along $S^1$ is already
   bounded we get
   \begin{equation*}
      \int_{S^1} \abs{\ubar{\nabla}^c ( r^2 \ubar{\nabla}_c \phi)} \le C
   \end{equation*}
   uniformly for each $t \in \,]t_1,0]$. Note, that we use the wave map
   $\phi=(V,W)$ defined in \ref{ss.field2+2} in order to abbreviate
   some formulas only.

   On the other hand, we can interpret the integrated quantity as a
   divergence, a fact we will take advantage of to get rid of the integral
   sign. To this end consider a point $(t,x)$ in the quotient manifold $B$,
   with $t \in \,]t_1,0[$. We need estimates uniformly in $t$ for $t$
   approaching $t_1$. We define the (upside down) characteristic triangle
   $T$ in $B$, by its counterclockwise oriented boundary 
   $\d T = \gamma_+ + \gamma_0 + \gamma_-$, where 
   $\gamma_{\pm}$ denote the characteristic curves of \eqref{e.fieldVWa},
   with $\dot{\gamma}_{\pm} = k_{\pm} = m \pm n$ connecting $(t,x)$ with
   $(0,x_{\pm})$ and $\gamma_0$ is the curve in the hypersurface $\{t=0\}$
   from $(0,x_+)$, to $(0,x_-)$ with $\dot{\gamma}_0 = -m$. Define  
   $v \in T^{\perp}F$ by $v^a =  r^2 \ubar{\nabla}^a \phi$, then
   the corresponding 1-form $\omega$ reads
   $\omega = \iota(v)\ubar{\Omega}$, where $\ubar{\Omega}$ denotes the
   volume form with respect to $\ubar{g}$. Expressing tensor components
   with respect to the projected orthonormal frame
   $\{\ubar{\sigma}_a=q_a\}$ (see \ref{ss.2+2}) we have 
   $v^a = r^2 \left(\begin{smallmatrix} 
                 -\ubar{\nabla}_0\phi \\ \ubar{\nabla}_1\phi  
               \end{smallmatrix}\right)$, 
   and 
   $\omega=-r^2 \left(m(\phi)\ubar{\sigma}^0 + n(\phi)\ubar{\sigma}^1\right)$.
   The pullback of $\omega$ along $\d T$ has the three parts
   $\gamma_{\pm}^*\omega = \mp r^2 k_{\pm}(\phi\circ\gamma_{\pm}(u))\,du$,
   $\gamma_0^*\omega = r^2 n(\phi(x)) A\,dx$. Thus, setting
   $\tau:=\sqrt{2}t$ 
   \begin{equation*}\begin{split}
      \int_{\d T}\omega &
        =   \int_{\tau}^0 -r^2 k_+(\phi(\gamma_+(u)))\,du 
          + \int_{x_+}^{x_-} n(\phi(x)) A\,dx
          + \int_0^{\tau} +r^2 k_-(\phi(\gamma_-(u)))\,du \\ &
        = - \int_{\tau}^0 r^2 k_+(\phi(\gamma_+(u)))\,du
          - \int_{\tau}^0 r^2 k_-(\phi(\gamma_-(u)))\,du
          + \int_{x_+}^{x_-} n(\phi(x)) A\,dx \\ &
        = -r^2\phi\Big|_{\gamma_+(\tau)}^{\gamma_+(0)}
          -r^2\phi\Big|_{\gamma_-(\tau)}^{\gamma_-(0)}
          + \int_{\tau}^0 k_+(r^2)\phi
          + \int_{\tau}^0 k_-(r^2)\phi
          + \int_{x_+}^{x_-} n(\phi(x)) A\,dx
   \end{split}\end{equation*}
   where integration by parts has been carried out in the last step. Note,
   that the last term is bounded, since that integration takes place on the
   line $t=0$ in $B$, where $\phi$ is smooth.
   In addition $k_{\pm}(r^2)$ is bounded everywhere by proposition
   \ref{prop.first_estimates2}. Further, the first terms can be
   evaluated to  
   $-r^2\phi\big|_{\gamma_+(\tau)}^{\gamma_+(0)}
    -r^2\phi\big|_{\gamma_-(\tau)}^{\gamma_-(0)} 
    = 2(r^2\phi)(t,x) - (r^2\phi)(0,x_+) - (r^2\phi)(0,x_-)$, where
   the last two terms live on the line $t=0$, and therefore are
   bounded.\zeile
   Stokes' theorem now applies to the present situation, 
   $\int_{\d T}\omega = \int_T dw = \int_T (\div v)\,\ubar{\Omega}$. Putting
   all this together we get an estimate
   \begin{equation*}
      \lnorm{r^2\phi(t)}{\infty} \le C
         \left( 
           1 + \int_t^0 \lnorm{r^2\phi(s)}{\infty}\,ds
             + \int_t^0 
               \int_{S^1}\abs{\ubar{\nabla}^c(r^2\ubar{\nabla}_c\phi)} \,ds
         \right)
   \end{equation*}
   where the last term is already known to be bounded uniformly in
   $t$ by our previous analysis. Hence we can apply Gronwall's inequality
   and the bounds for $r$, 
   $r^{-1}$ obtained in proposition \ref{prop.first_estimates2} complete
   our argument, arriving finally at $\abs{V},\abs{W} \le C$ uniformly on
   $]t_1,0[$. 
\end{proof}

Unfortunately, there is still a lack of control over the components
$\kappa$ of $k$ (respectively its traceless part, since
$(\tr\kappa)\,\ubar{g}=(H-K)\ubar{g} \le C$). It turns out, that the
necessary bounds for the first derivatives of $\phi$ depend on a bound 
for the matter quantities and vice versa. The analysis of this
dependence will be performed in the next section, where we have to
take a specific matter model into account.
%
%
%
%
%
%
%
%
%
%
\subsection{Higher order estimates}
\subsubsection{First order estimates}\label{ss.1order}
We start with the derivation of first order estimates for the metric
coefficients $\phi=(V,W)$. To keep things simple, we consider
collisionless matter, as described by the Einstein Vlasov system. 

The particle density $f$ on the mass shell is governed by the geodesic
spray $X$ with 
$X = v^{\alpha}\viere_{\alpha} 
   - \viergamma_{\mu\nu}^\kappa v^{\mu}v^{\nu} \tfrac{\d}{\d v^{\kappa}}$,
and the killing vector fields $\d_A$ define two conserved
quantities $g(X,\d_A)$, which bound $v^A$. To bound $v^1$ we consider the 
characteristics of the Vlasov equation (compare the explicit formulas in
\cite{h3}): 
\begin{equation*}\begin{split}
   \frac{dv^1}{ds} &
     = -\left( e_1(N)v^0 
             + N(-k_{rs}{e_1}^r{e_j}^s+\viergamma_{0j}^1)v^j
             + N\gamma_{jk}^1 v^jv^k/v^0
        \right) \\ &
     \le C \left( 1 + Q_1(s) + v^1(s) \right)
   \komma
\end{split}\end{equation*}
where $Q_1 := \lnorm{\ubar{D}\phi}{\infty}$, $\ubar{D} := (\d_t,\d_x)$
and we used the fact, that in view of the boundedness of $v^A$, the
vanishing of $\viergamma_{01}^1$ and the  
special form of the non-vanishing  rotation coefficients no term
involving the product $Q_1v^1$ occurs. Thus we find for the quantity
$\bar{P}_f(s) = \{\sup\abs{v} \,|\, v \in \supp f \}$
(which measures the matter quantities, compare \cite{h3}) the Gronwall-like
estimate 
\begin{equation*}
   1 + \bar{P}_f(t) \le C 
     \left( 1 + \bar{P}_f(0) + \int_t^0 (\bar{P}_f(s) + Q_1(s))\,ds \right)
\end{equation*}

This inequality shows, that we indeed have to estimate the matter
quantities together with the second fundamental form $\kappa$. The
next step consists in finding a complementary inequality for $Q_1$,
which yields in combination with the inequality just obtained a true
Gronwall estimate for $1+\bar{P}_f+Q_1$.\zeile
We need again the field equations for $V$ and $W$, but now it is more
convenient to analyse the equations in the 'wave map form'
\eqref{e.fieldVWb}. First consider the left-hand side of the field
equation. We want to express this derivative operator 
in terms of the characteristic vector fields $k_{\pm}=m \pm n$ tangent to
the characteristic curves $\gamma_{\pm}$ introduced in the previous section
(in the proof of proposition \ref{prop.second_estimates}). In the quotient
manifold $(B,\ubar{g})$ one calculates the non-vanishing Ricci rotation
coefficients to 
\begin{gather*}
   \ubar{\gamma}_{01}^0=\ubar{\gamma}_{00}^1=(AN)^{-1}N' \\
   \ubar{\gamma}_{11}^0=\ubar{\gamma}_{01}^1=(AN)^{-1}(\dot{A}-(A\nu^1)')
   \komma
\end{gather*}
thus $\ubar{\nabla}_{\ubar{e}_a}\ubar{e}_b$ is bounded. The
characteristics $k_{\pm}$ expressed in the coordinates of $B$ are simple
linear combinations of the $\ubar{e}_a$, hence 
$\ubar{\nabla}_{k_{\pm}}k_{\pm}$ is also bounded in $B$. We will use this
fact to estimate commutators such as $[m,n]$ or $[k_+,k_-]$. Transforming
from the orthonormal frame $\{\ubar{e}_a\}$ in $(B,\ubar{g})$ to the frame 
$\{k_{\pm}\}$, we can write the left-hand side of \eqref{e.fieldVWb} as
\begin{equation*}
   \ubar{\nabla}^c \ubar{\nabla}_c \phi^K
   + \hat{\Gamma}_{IJ}^K \ubar{\nabla}^c \phi^I \ubar{\nabla}_c \phi^J
   = \ubar{\nabla}_{k_+} \left(k_-(\phi^K)\right)
   + \hat{\Gamma}_{IJ}^K \,k_+(\phi^I) k_-(\phi^J)
   + [k_+,k_-](\phi^K)
\end{equation*}
and there is an analogous equation for $k_+$ and $k_-$ interchanged.\zeile
Let us turn now to the right-hand side of \eqref{e.fieldVWb}. The terms not
known to be bounded are of the form $k_{\pm}(\phi)$ and $T_{AB}$. 
Fortunately we have control over the matter term, since the bounds on $V,W$
and $v^A$ justify the inequality $T_{AB} \le C\,\bar{P}_f$ and we obtain
\begin{equation*}
   \ubar{\nabla}_{k_+} \left(k_-(\phi^K)\right)
   + \hat{\Gamma}_{IJ}^K \,k_+(\phi^I) k_-(\phi^J)
     \le C \Bigl( 1 + k_{\pm}(\phi) + \bar{P}_f \Bigr)
   \punkt
\end{equation*}
The left-hand side is still non-linear in the first
derivatives of $\phi$, so this inequality is not in the appropriate form to
apply some kind of Gronwall estimate. But we can overcome this difficulty
by adapting an observation of Gu (\cite{gu}, see also \cite{r} for
explanation), who considered wave maps
defined on two dimensional Minkowski space. We already performed the first
step, consisting of a transformation to characteristic coordinates. Now we
define the vector field $\hat{k}_{\pm}$ over the map $\phi$,
\begin{equation*}
   \hat{k}_{\pm} := \phi_*(k_{\pm})
                  = k_{\pm}(\phi)
                  = k_{\pm}(\phi^K)\frac{\d}{\d \phi^K}
                  = k_{\pm}(V)\frac{\d}{\d V} 
                  + k_{\pm}(W)\frac{\d}{\d W}
   \komma
\end{equation*}
thus bounding the length of $\hat{k}_{\pm}$ accomplishes the estimate
of $Q_1$. The vector field $\hat{k}_{\pm}$ over $\phi$ is a section in
the bundle $\phi^*(TH^2)$ over $B$ and the 
covariant derivative operator $\phi^*\hat{\nabla}$ over $\phi$ acts like
$(\phi^*\hat{\nabla})_v \hat{X} 
   = v(\hat{w}^K)\tfrac{\d}{\d \phi^K} 
   + \hat{w}^K \hat{\nabla}_{\phi_*v}\tfrac{\d}{\d \phi^K}$
for every vector $v \in TB$ and every 
$\hat{X}=\hat{w}^K \tfrac{\d}{d \phi^K}$ with $w^K:B \rightarrow H^2$.\zeile
Evaluating this expression with $v={k_+}$ and $\hat{X}=\hat{k}_-$ gives
\begin{equation*}
   (\phi^*\hat{\nabla})_{k_+} \hat{k}_- 
     = k_+(k_-(\phi^K))\frac{\d}{\d \phi^K}
     + \hat{\Gamma}_{IJ}^K\,k_+(\phi^I)k_-(\phi^J)\frac{\d}{\d \phi^K}
\end{equation*}
(and analogously for $k_{\pm}$ interchanged). The components in this
equation are similar to the left-hand side of the inequality already
obtained for the field equations. The remaining term involves 
$k_{\pm}(\phi)$ times the connection coefficients in $B$ with respect to the
frame $\{k_{\pm}\}$, which we know to be bounded. Thus we get for the field
equations the inequality
\begin{equation*}
   \abs{(\phi^*\hat{\nabla})_{k_+} \hat{k}_-} 
     \le C \Bigl( 1 + k_{\pm}(\phi) + \bar{P}_f \Bigr)
   \komma
\end{equation*}
expressing an estimate about the growth of $\hat{k}_{\pm}$ during the
transport along the characteristic curve $\gamma_{\mp}$. This 
inequality is in the appropriate form
for a Gronwall-like estimate: Taking the maximum for each fixed $t$ allows
us, to combine the inequalities for $\abs{\hat{k}_+}$ and
$\abs{\hat{k}_-}$, replaced collectively in terms of $Q_1$. Hence
\begin{equation*}
   Q_1(t) \le C \left( 
     1 + Q_1(0) + \int_t^0 (\bar{P}_f(s) + Q_1(s))\,ds 
                \right)
   \komma
\end{equation*}
where $\gamma_{\pm}$ has been reparametrized in terms of $t$.
Combining this with the inequality for $\bar{P}_f$ we arrive at
\begin{equation*}
   1 + \bar{P}_f(t) + Q_1(t) \le
     C \left(
         1 + \bar{P}_f(0) + Q_1(0) 
         + \int_t^0 (1 + \bar{P}_f(s) + Q_1(s))\,ds
       \right)
   \punkt
\end{equation*}
Performing a Gronwall argument we have proven
\begin{proposition}\label{prop.1order_estimates}\ein
   Let $(M,g,f)$ be a globally hyperbolic, spatially compact solution of
   the Einstein-Vlasov system with local $U(1) \times U(1)$ symmetry, which
   possesses a symmetric Cauchy surface $\Sigma$ with strictly negative mean
   curvature. The PMC time coordinate $t$ ranges in $]t_1,t_2[$ with
   $\Sigma=\{t=0\}$ and $H$ decreases with decreasing $t$.\zeile
   Then we get uniform bounds for  
   \begin{equation*}
      \abs{\dot{V}}, \abs{\dot{W}}, \abs{V'}, \abs{W'}, \rho
   \end{equation*}
   on $]t_1,0]$.
\end{proposition}
\subsubsection{Second order estimates}
We still have to do one further step before it will be possible to apply
some iteration scheme. We need to establish bounds for the second
derivatives of $\phi=(V,W)$ together with bounds for the first derivatives
of the matter variables. Again it turns out, that it is not possible to get
separate estimates for these quantities. 

First we differentiate the field equation \eqref{e.fieldVWb} and obtain an
equation of the form 
\begin{equation*}
   k_{\pm} \left(k_{\mp}(\d_x \phi^K)\right)
   + 2 \hat{\Gamma}_{IJ}^K \, k_{\pm}(\phi^I) k_{\mp}(\d_x \phi^J) 
   = [...] \d_x T_{AB} + \text{lower order terms}
   \komma
\end{equation*}
where $[...]$ is an abbreviation for some term involving already bounded
quantities, in particular we see by a quick look at the 3+1--field
equations, whose right-hand sides are now bounded, that for the Ricci
rotation coefficients $\d_x \ubar{\gamma} \le C$ holds. 
Note, that the differentiation kills the nonlinearity in the equation,
but we have to deal with first order derivatives of the matter
variables instead. These quantities cause serious trouble, because if we
differentiate the Vlasov equation, terms involving the second
derivatives of $\phi$ times first derivatives of $f$ come up, thus there is
no direct Gronwall argument possible. To attack this difficulty we have to
combine the equations, using an idea of \cite{gs}.\zeile
First we integrate the equation along the characteristics $\gamma_{\pm}$ to
get an integral inequality. In order to apply a Gronwall argument all what
remains to do is to bound the term 
$\int_{\gamma_{\pm}} \d_x T_{AB}$. 
Then we express $T_{AB}$ in terms of its components with
respect to the orthonormal frame, producing merely some already bounded
quantities and the frame components of the energy-momentum tensor
for Vlasov-type matter look like 
$\int_{\gamma_{\pm}} \left( \int v_A v_B/v^0\, \d_x f \,dv\right)$.
Finally we represent $\d_x$ by a linear combination of $k_{\pm}$ and 
$\ubar{X} : = \d_t + (NA^{-1}v^1/v^0-\nu^1)\d_x$, which is the
part of the 
characteristic vector field for the Vlasov equation, lying in $B$. The
transformation to this basis is obviously bounded and we can proceed as
follows. The part involving $k_{\pm}$ permits a direct application of the
integration by parts rule, which contributes something bounded by
proposition \ref{prop.1order_estimates}. The remaining
part can be treated in a similar fashion, after inserting the Vlasov
equation into $\ubar{X}(f)$. This yields terms involving only first order
derivations of $\phi$ (arising from the $\viergamma$'s) times 
$\tfrac{\d f}{\d v}$, and again we can perfom an integration by parts with
respect to the velocity integral. The result consists in terms
already bounded and we are through. Applying now Gronwall's inequality we
arrive at a bound for $\lnorm{\d_x\ubar{D}\phi}{\infty}$ and inserting
this result into the field equation \eqref{e.fieldVWb} we get
$Q_2 := \lnorm{\ubar{D}^2\phi}{\infty}$ bounded.\zeile
An immediate
consequence is, that we also have established a bound for the first
derivatives of $f$. Differentiating the Vlasov equation with respect to $x$
or $v$ yields an equation for $\d_x f$ or $\d_v f$ respectively, with
bounded characteristics and an inhomogeneous term, consisting of
$(\viere_{\mu})^{\alpha}$, $\viergamma_{\mu\nu}^\kappa$ and their
derivatives with respect to $x$. These terms are either bounded by what has
been said above or by inspection of the field
equations \eqref{e.ch2}--\eqref{e.sB2}. Having bounded the spatial and
velocity derivatives of $f$, the structure of the Vlasov equation bounds
immediately $\d_t f$ and we have proven
\begin{proposition}\label{prop.2order_estimates}\ein
   Under the hypotheses of proposition \ref{prop.1order_estimates} we
   get uniform bounds for 
   \begin{equation*}
      \abs{\ddot{V}}, \abs{\dot{V}'}, \abs{V''},
      \abs{\ddot{W}}, \abs{\dot{W}'}, \abs{W''}
      \quad\text{and}\quad
      \abs{\dot{\rho}}, \abs{\rho'}
   \end{equation*}
   on $]t_1,0]$.
\end{proposition}
\subsubsection{The iteration scheme}
Following the analysis of part I we need some
additional framework. 
{\samepage
\begin{definition}\label{def.collect2}\ein\nopagebreak\vspace{-0.4cm}
   \begin{equation*}
      \mathcal{F} := \left(
              A,a,V,W,N,\nu^1,\nu^B,H,K,\eta_B
           \right)
   \end{equation*}
   \begin{equation*}
      \Phi := \left(
                 \rho,\leftsuper{\phi}\rho
              \right)
   \end{equation*}
\end{definition}
}
It turns out, that what we have done in the previous subsections are the
first steps towards the matter regularity property of part I. Here we will
define this property adapted to the present situation with the 
notational conventions introduced as follows: 
$\alpha$ denotes a multi index for derivatives in $(B,\ubar{g})$ and
$\ubar{D}:=(\d_t,\d_x)$. Then the matter regularity property reads
\begin{equation}\label{e.matter_regularity2}
   \abs{\ubar{D}^{\alpha}{\cal F}} \le C 
   \qquad\Longrightarrow\qquad
   \abs{\ubar{D}^{\alpha}\Phi} \le C 
\end{equation}
and we state
\begin{proposition}\label{prop.vmatter_regularity}\ein
   Under the hypotheses of proposition \ref{prop.1order_estimates}
   \eqref{e.matter_regularity2} holds for all $\alpha$ uniformly
   on $]t_1,0]$.
\end{proposition}
\begin{proof}\zeile
   We prove the statement by induction with respect to $\abs{\alpha}$. The
   induction hypotheses for $\alpha=0$ is contained in the statements of
   the propositions \ref{prop.1order_estimates} and
   \ref{prop.2order_estimates}. To proceed further let us assume 
   $\abs{\ubar{D}^{\alpha}{\cal F}} \le C$ for some $\alpha$ with
   $\abs{\alpha} =: p \ge 1$. Then we can assume 
   $\abs{\ubar{D}^{p-1} \Phi} \le C$ by induction. We have to show 
   $\abs{\ubar{D}^p \Phi} \le C$, or extending our previous notation,
   $P_p := \lnorm{\ubar{D}^p f}{\infty} \le C$ and
   $Q_{p+1} := \lnorm{\ubar{D}^{p+1} \phi}{\infty} \le C$.

   Following the analysis of the previous subsection we set $q := p-1$ 
   and differentiate the
   field equations \eqref{e.fieldVWb} with $\d_x\ubar{D}^q$, which
   yields schematically 
   \begin{equation*}\begin{split}
      k_{\pm}\bigl(k_{\mp}(\d_x\ubar{D}^{q}\phi)\bigr) = &
        [\d_x\ubar{D}^q\ubar{\gamma}, \d_x\ubar{D}^{p}r, \d_x\ubar{D}^q\phi] 
        \bigl( k_{\mp}\d_x\ubar{D}^{q}\phi \bigr) \\ & 
      + [\ubar{D}^{q}\phi]\, \d_x\ubar{D}^q T_{AB}
      + [\d_x\ubar{D}^q\phi, \d_x\ubar{D}^q \eta, \ubar{D}^q T_{AB}]
      \komma
   \end{split}\end{equation*}
   where quantities in square brackets abbreviates some expressions
   formed by them, whose detailed structure is not important for our
   analysis here. We want to perform the same kind of argument as we have
   done in the proof of the second order estimates. In order to do this we
   must bound the quantities in the square brackets:
   \begin{itemize}
   \item The terms in the second and third square bracket are bounded
      by the induction hypotheses.
   \item For the first square bracket we can proceed as
      follows. $\d_x\ubar{D}^q\phi$ is already bounded. For 
      $\d_x\ubar{D}^{p}r$ we have to estimate 
      $\d_x\ubar{D}^p A=\d_x^2\ubar{D}^q A$. This
      can be done by applying $\ubar{D}^q$ to \eqref{e.ch2} in a
      strictly analogous manner 
      as in the first item in the proof of lemma 3.6 in \cite{h3},
      applied to equation \eqref{e.ch2}. The
      definition of $\ubar{\gamma}$ in \ref{ss.1order} shows, that the
      terms in $\d_x\ubar{D}^q \ubar{\gamma}$ not
      already bounded are $\d_x^2\ubar{D}^q \nu^1$, 
      $\d_x^2\ubar{D}^q N$ and $\d_x\d_t\ubar{D}^q A$. 
      Applying $\ubar{D}^p$ to \eqref{e.s2} gives a bound for the
      first quantity. Using $\ubar{D}^q$ on \eqref{e.l2}, then the
      argument in the fourth step in the proof of lemma 3.6 in \cite{h3},
      applied to 
      \eqref{e.l2} bounds the second quantity. Now all quantities
      appearing on the right-hand side of ($\d_x\ubar{D}^q$ applied
      to) \eqref{e.e12} are bounded, and this bounds 
      $\d_t\d_x\ubar{D}^q A$.
   \end{itemize}
   Turning now to the Vlasov equation and applying $\d_x\ubar{D}^q$ yields
   an inhomogeneous equation for $\d_x\ubar{D}^q f$ with the same
   characteristics. Thus we can apply the same trick as in the
   previous subsection, substituting $\d_x$ by $k_{\pm}$ and
   $\ubar{X}$. Then again we are concerned with integration by parts,
   which yields only bounded terms, and the inhomogeneous term of the
   differentiated Vlasov equation, which is also bounded by induction
   hypotheses. All together we can apply Gronwall's inequality to find
   $\lnorm{\d_x\ubar{D}^p\phi}{\infty} \le C$, and immediately 
   $Q_{p+1} \le C$ by inserting the spatial bounds into the field
   equation. The Vlasov equation for $\d_x\ubar{D}^q f$ now bounds 
   $\d_x\ubar{D}^q f$ and analogously for $\d_v\ubar{D}^q f$, which
   automatically bounds $\d_t\ubar{D}^q f$ by inserting the spatial
   and velocity bounds into the differentiated Vlasov equation. Thus
   we have also bounded $P_p$, which completes the proof of the
   proposition. 
\end{proof}

To extend the local in time PMC foliation, our aim is to find
uniform $C^{\infty}$ bounds of all geometric and matter
quantities, when the PMC time $t$ approaches $t_1$. Property
\eqref{e.matter_regularity2} shows, that it is 
enough to bound the geometric quantities $\ubar{D}^{\alpha}{\cal F}$
for all $\alpha$ as long as $t$ or respectively the mean curvature remains
finite . Therefore all we need is the analogue of the lemmas 
3.6 and 3.7 in \cite{h3}, starting
with the propositions \ref{prop.first_estimates2},
\ref{prop.second_estimates}, \ref{prop.1order_estimates} and
\ref{prop.2order_estimates}.\zeile 
The first part is straightforward, given an arbitrary multi index
$\alpha$ and 
$\abs{\ubar{D}^{\alpha}{\cal F}} \le C$ we find 
$\abs{\d_x\ubar{D}^{\alpha}{\cal F}} \le C$ by inspection of the
relevant (differentiated) field equations
\eqref{e.ch2}--\eqref{e.sB2} (compare the arguments given in the proof 
of lemma 3.6 in \cite{h3} where necessary), with the terms 
on the right-hand sides bounded by property
\eqref{e.matter_regularity2}. Moreover, proposition
\ref{prop.vmatter_regularity} provides not only bounds for the spatial
derivatives, but also for the time derivatives of
$\phi$ by the definition of $\leftsuper{\phi}\rho$. This allows us to
proceed as 
follows: First we bound, starting from $\ubar{D}^{\alpha}{\cal F}$ all 
spatial derivatives $\d_x^k\ubar{D}^{\alpha}{\cal F}$, $k=1,2,\dots$
Then we bound, starting successively from  
$\d_x^k\ubar{D}^{\alpha}{\cal F}$, $k=1,2,\dots$ the quantities
$\ubar{D}^{\beta}{\cal F}$ for all multi indices $\beta$ with 
$\abs{\beta}=\abs{\alpha}+k$, redistributing the spatial derivatives
into time derivatives. This procedure will successively bound all
derivatives of ${\cal F}$, taking advantage of the fact, that in each
step all lower order derivatives with the same order of time
derivatives have been bounded as well as at least one more spatial
order derivative in the step with one order less in time derivatives.
In a more compact formulation, we have to show the boundedness of 
$\d_t\ubar{D}^{\alpha}{\cal F}$, having already bounds for 
$\d_x\ubar{D}^{\alpha}{\cal F}$ and 
$\ubar{D}^{\alpha}{\cal F}$. 

We accomplish this step for each member of ${\cal F}$, by again
considering some of the (differentiated by $\ubar{D}^{\alpha}$) 
3+1--field equations. We see immediately from \eqref{e.e12}
and \eqref{e.e22}, that we have bounds for 
$\d_t\ubar{D}^{\alpha} A$, $\d_t\ubar{D}^{\alpha} a$. The bounds for 
$\d_t\ubar{D}^{\alpha}$ applied on $H$, $K$ and $\eta$ are
straightforward, too. Of course, $\d_t\ubar{D}^{\alpha}\phi$ is
already bounded by property \eqref{e.matter_regularity2}. Moreover we
can strengthen the regularity. Since we have already a bound for 
$\d_x\ubar{D}^{\alpha}{\cal F}$ proposition
\ref{prop.vmatter_regularity} provides us with bounds for
$\d_x\ubar{D}^{\alpha}$ applied to $\rho$ and
$\ubar{D}\phi$. Applying $\d_x\ubar{D}^{\alpha}$ to the field equation
\eqref{e.fieldVWb} we first get a bound for
$\d_x\ubar{D}^{\alpha}\ubar{D}\phi$ and then for 
$\d_t^2\ubar{D}^{\alpha}\phi$ by inserting the first result into the
field equation. This in turn used together with the bounds for the
differentiated matter variable in the {differentiated} Vlasov
equation, bounds $\d_t\ubar{D}^{\alpha}\rho$.\zeile
Now we turn to the analysis of the differentiated lapse equation. 
First we see, that the bound for $\d_t\ubar{D}^{\alpha} \nu$ follows
from the bound of $\d_t\ubar{D}^{\alpha} N$. For the latter one,
we follow the argument given in the corresponding place in the
proof of lemma 3.7 in \cite{h3}, where all that is needed, has already
been bounded by the arguments just given here and we are done.

Therefore we end up with
{\samepage
\begin{theorem}\label{thm.ev2}\ein
   Let $(M,g,f)$ be a globally hyperbolic, spatially compact solution of
   the Einstein-Vlasov system with local $U(1) \times U(1)$ symmetry, which
   possesses a symmetric Cauchy surface $\Sigma$ with strictly negative mean
   curvature $H \le \bar{H} < 0$ and $H=\bar{H}$ somewhere on
   $\Sigma$.\zeile
   Then all of the past of $\Sigma$ admits a PMC foliation
   $\{S_t\}$, where $t$ takes all values in the interval
   $\,]\!-\infty,0]$ and $H$ takes all values in
   $\,]\!-\infty,\bar{H}]$.
\end{theorem}
}
\subsubsection{Improving the result}
Here we will try to do the same construction as in the corresponding place
in \cite{h3}, to get
rid of the restriction concerning the fixed sign of the mean curvature $H$
on the Cauchy surfaces. So we assume $M$ to be non-flat and denote by
$\Sigma$ an arbitrary symmetric Cauchy surface in $M$.
In order to follow the steps performed in the plane symmetric case in
part I, remember first from \ref{ss.expanding2}, that the
spacetime is expanding. Note further, that the Einstein-Vlasov system
fulfills the dominant and strong energy condition as well as the
non-negative pressures condition. 

Then we find for the past domain of dependence $D^-(\Sigma)$
\begin{itemize}
\item Since the spacetime is expanding, $dr$ is future pointing, thus $r$
   is bounded in $D^-(\Sigma)$. By assumption, $\mass^{-1}$ is bounded on
   the compact surface $\Sigma$. Unfortunately, unlike in the plane
   symmetric case we cannot conclude from the non-negative pressures
   condition and the fact that $dr$ is future pointing, that $d\mass$ is past
   pointing, since in equation \eqref{e.massflux2} for the mass flux the
   term involving $\eta$ contributes with the wrong sign (luckily, the
   contribution of the energy momentum tensor $\leftsuper{\phi}\ubar{T}$
   \eqref{e.Tphi} for the wave map $\phi$ does not cause any trouble, due
   to the energy conditions automatically fulfilled by
   $\leftsuper{\phi}\ubar{T}$).\zeile
   To overcome this difficulty we impose the Gowdy--type symmetry condition,
   thus $\eta$ vanishes identically in the spacetime (and by the gauge
   condition $\nu^B(t,0)=0$ this is also true for $\obar{\nu}$ in view of
   \eqref{e.sB2}). Then we conclude, that $\mass^{-1}$ is bounded in
   $D^-(\Sigma)$ as desired. 
   
   Now we want to apply theorem 2.1 in \cite{b} to establish a bound for
   the length of all timelike curves in $D^-(\Sigma)$. For this we have to
   adapt the proof a little bit. Inspection of the proof shows, that the
   argument relies on the inequality 
   $\ddot{r} \le -\tfrac{\mass}{r^2}$ 
   satisfied by the area radius along timelike geodesics, where the dot
   denotes differentiation along the curve. But looking at the field
   equation \eqref{e.fieldr2} for the area radius easily establishes this
   relation (since $\eta$ vanishes) and we are through.
\item The basic estimate \eqref{e.mom2} together with 
   \eqref{e.mass2} shows, that $r^{-1}$ and $\mass$ are bounded.
\item Lemma 2 in \cite{br} applies and we get bounds for the volume
   and its inverse for any Cauchy surface in $D^-(\Sigma)$.
\end{itemize}
The volume $V(t)$ for the leaf $S_t$ of the PMC foliation is given by 
$V(t)=(4\pi)^2 a^{-1} \int_{S^1}r^3$ and a closer look at the
estimates given in the corresponding place in \cite{h3} shows, that the
arguments apply literally. 
This relies on the facts, that on the one hand the additional equations
\eqref{e.fieldVWb} are unaffected by the 
reparametrization of the foliation, thus the estimates for $\phi$ and
its derivatives done in the previous subsections hold. On the other
hand the construction in part I is based mainly on the
structures introduced by the matter regularity property and the lemmas 
3.6 and 3.7, a program we adapted successfully to the more
general situation here. 

So we finally arrive at
\begin{theorem} \label{thm.pmcu1}  \ein
   Let $(M,g,f)$ be a globally hyperbolic, spatially compact solution of the
   Einstein-Vlasov system with Gowdy--type local $U(1) \times U(1)$ symmetry
   and $\Sigma$ be a symmetric Cauchy surface.\zeile
   If $(M,g)$ is non-flat then we can foliate all of
   the past of $\Sigma$ by  
   PMC hypersurfaces, where the time function takes on all values in the
   interval $]-\!\infty,0]$ and the mean
   curvature of the leaves tends uniformly to $-\infty$ for 
   $t \rightarrow -\infty$.
\end{theorem}
Using the PMC leaves as barrier surfaces we get the
\begin{corollary}\label{cor.cmcu1}\ein
   In the situation of theorem \ref{thm.pmcu1} $D^-(\Sigma)$ possesses a
   CMC Cauchy surface for each value of the mean curvature in
   $]-\!\infty,\min_{\Sigma}H[$. 
\end{corollary}
%
%
%
%
%

%
%
%
%
%
%
\section{Conclusion and outlook}
We have seen in this second part, that the program initiated in part I has
been successfully extended to the more general case of local 
$U(1) \times U(1)$ symmetric spacetimes in close analogy to the plane
symmetric spacetimes. Therefore all what has been mentioned in
the corresponding place in \cite{h3} applies.

A corresponding analysis for spacetimes with local $U(1) \times U(1)$
symmetry has been independently performed in the
work \cite{an} of Andr\'easson. His construction leads to stronger results
(as the work \cite{re} of Rein in the surface symmetric case of part I), but
the time functions used are defined in terms of the symmetry. Thus it is
not clear, how to generalize them. In view of Andr\'easson's work, the
present analysis can be seen as a suggestion pointing in a slightly
different direction. 

In comparison with the spacetimes considered in part one the control of the
momenta of the Vlasov particles turned out to be more complicated,
in particular the coupling to gravitational waves 
required second order estimates before the iteration procedure could be
performed. In this process we took advantage from the wave-map structure
of the dynamical part of the geometry driven by the simple
form of the Vlasov equation.\zeile
Nevertheless one can hope, that the approach may be generalized to other
matter models and geometries, since the obstructions seem to be of technical
nature only.

\vspace{1em}\zeile
\textbf{Acknowledgements:}
I want to thank Alan Rendall, who guided me through these parts of my
work, which depend heavily on his previous developments.
%
%
%
%
%


%
%
%
%
%
%
\pagebreak

%
%
%
%
%

%
%
%
%
%
\end{document}